\RequirePackage{silence}
\documentclass[%
a4paper,
prd,
twocolumn,
reprint,
superscriptaddress,
preprintnumbers,
nofootinbib,
nobibnotes,
amsmath,
amssymb,
aps,
floatfix
]{revtex4-2}
\usepackage[english]{babel}
\usepackage{amsfonts,mathrsfs,physics}
\usepackage{color,graphicx,xcolor}
\usepackage[bookmarks=true,
pdfnewwindow=true,
colorlinks=true,
linkcolor=xlinkcolor, citecolor=xlinkcolor, filecolor=xlinkcolor, urlcolor=xlinkcolor,
final=true
]{hyperref}
\usepackage{dcolumn,multirow}
\definecolor{xlinkcolor}{cmyk}{1,1,0,0}
\newcommand{\appsection}[1]{
  \refstepcounter{section}
  \section*{APPENDIX \thesection: #1}
  \addcontentsline{toc}{section}{APPENDIX \thesection: #1}
}

\makeatletter
\AtBeginDocument{%
  \renewcommand{\bibsection}{%
    \par
    \onecolumngrid@push
    \vspace*{\baselineskip}
    \begingroup
      \phantomsection
      \addcontentsline{toc}{section}{\protect\numberline{}\refname}%
      \noindent\rule{\textwidth}{0.5pt}%
    \endgroup
    \nobreak\@nobreaktrue
    \addvspace{19\p@}%
    \par
    \onecolumngrid@pop
  }%
}
\makeatother

\begin{document}

\preprint{INR-TH-2026-007}

\title[Constraining Lorentz invariance violation from the depth of air-shower maximum]{Constraining Lorentz invariance violation from the depth of air-shower maximum}
\author{Nickolay\,S.\,Martynenko}
\thanks{Corresponding author, e-mail: martynenko@inr.ac.ru}
\affiliation{Lomonosov Moscow State University, 1-2 Leninskie Gory, Moscow 119991, Russia}
\affiliation{Institute for Nuclear
Research of the Russian Academy of Sciences, 60th October Anniversary Prospect 7a, Moscow 117312, Russia}
\author{Grigory\,I.\,Rubtsov}
\affiliation{Institute for Nuclear
Research of the Russian Academy of Sciences, 60th October Anniversary Prospect 7a, Moscow 117312, Russia}
\author{Petr\,S.\,Satunin}
\affiliation{Institute for Nuclear
Research of the Russian Academy of Sciences, 60th October Anniversary Prospect 7a, Moscow 117312, Russia}
\affiliation{Lomonosov Moscow State University, 1-2 Leninskie Gory, Moscow 119991, Russia}
\affiliation{Branch of Lomonosov Moscow State University in Sarov, Parkovaya 2, Sarov 607328, Russia}
\author{Andrey\,K.\,Sharofeev}
\affiliation{Lomonosov Moscow State University, 1-2 Leninskie Gory, Moscow 119991, Russia}
\affiliation{Institute for Nuclear
Research of the Russian Academy of Sciences, 60th October Anniversary Prospect 7a, Moscow 117312, Russia}
\author{Sergey\,V.\,Troitsky}
\affiliation{Institute for Nuclear
Research of the Russian Academy of Sciences, 60th October Anniversary Prospect 7a, Moscow 117312, Russia}
\affiliation{Lomonosov Moscow State University, 1-2 Leninskie Gory, Moscow 119991, Russia}
\date{
Submitted to \textit{Physical Review D} on August 6, 2026
}

\begin{abstract}
Hypothetical Lorentz invariance violation (LIV) remains strongly constrained but has not been excluded as a possible manifestation of new high-energy physics. In subluminal photon-sector LIV, suppressed Bethe--Heitler pair production modifies the electromagnetic component of extensive air showers. We present a simulation-based toy analysis of this effect using published Pierre Auger fluorescence-detector distributions of the reconstructed depth of air-shower maximum. The modified distributions of the depth of the shower maximum are folded with the Auger detector-response parametrizations and compared with the data using free composition fractions in each reconstructed-energy bin. Within this approach, the comparison gives a sensitivity to LIV compatible with that expected from ultra-high-energy photon-induced showers. We obtain \(M_{\rm LIV}\gtrsim 6.5\times 10^{18}\,\text{GeV}\) for the photon-sector LIV mass scale \(M_{\rm LIV}\) (\(95\%\) confidence level). This result should not be interpreted as a detector-level experimental limit because the detector response and reconstruction are simplified, and the result remains composition dependent. However, the approach presented here opens the way for constraining LIV in hadronic air showers with more detailed analyses.
\end{abstract}

\maketitle

\section{Introduction}
\label{sec:introduction}

Hypothetical Lorentz invariance violation (LIV) remains one of a few possible manifestations of new physics at high energies that has not yet been experimentally ruled out. Such effects arise naturally in several approaches to quantum gravity and are widely discussed as possible phenomenological signatures of physics beyond the Standard Model \cite{Addazi2022QuantumGravityReview, Colladay:1998fq}. Although existing astrophysical and laboratory constraints are already very strong \cite{Kostelecky:2008ts}, some LIV scenarios remain testable only at extreme energies. Ultra-high-energy cosmic rays may therefore serve as a useful probe, since their interactions in the atmosphere can reach regimes inaccessible to accelerator experiments~\cite{Jacobson2006HighEnergyLIVReview,MartinezHuerta2020AstroparticleLIV,Addazi2022QuantumGravityReview,Auger2022AstroparticleLIV,Auger2026MuonFluctuationsLIV}.

In this work, we consider a photon-sector LIV scenario with a subluminal photon group velocity. It results in modification of high-energy electromagnetic interactions and in suppression of Bethe--Heitler pair production by photons compared to the Lorentz-invariant (LI) case~\cite{BetheHeitler1934PairProduction,Vankov2002QEDFormationLength,Rubtsov2012LIVCrossSections}. In an electromagnetic cascade, this suppression delays the conversion of energetic photons into electron--positron pairs and changes the way the electromagnetic energy is deposited in the atmosphere. The effect can therefore influence not only rare photon-induced showers, but also the electromagnetic component of regular extensive air showers (EASs) induced by primary nuclei.

Photon-sector LIV has been constrained through several complementary channels. Observations of high-energy photons from astrophysical sources test modifications of photon propagation, shifts of pair-production thresholds, and possible changes in shower formation. Constraints have been obtained from the spectra of gamma-ray sources, from photons observed above tens or hundreds of TeV, and from the non-observation of expected fluxes of ultra-high-energy photons~\cite{Galaverni2008UHEPhotons,Lang2018UHECRPhotonLIV,Lang2019GammaRayLIV,Rubtsov2017TeVPhotonShowers,Satunin2019CrabLIV,Satunin2021TibetLHAASOLIV,Auger2022AstroparticleLIV}. These studies show that electromagnetic processes at extreme energies provide some of the most sensitive probes of subluminal photon-sector LIV.

EAS development provides a more direct way to search for LIV effects in particle interactions. Since the longitudinal profile of an EAS is shaped by the competition between hadronic particle production, neutral-pion decay, electromagnetic cascading, and energy losses in the atmosphere, modified electromagnetic processes can leave observable signatures in the depth of shower maximum, \(X_{\max}\), and in the shape of the shower profile. Early studies proposed to use the distribution of shower maxima as a test of Lorentz invariance, while later work developed this idea in specific photon-sector scenarios, including modified electromagnetic shower development, photon decay, and vacuum Cherenkov emission, and demonstrated that shower-maximum-related observables can provide competitive constraints~\cite{Antonov2001LongitudinalDevelopmentLIV,GagnonMoore2004UHECRLIV,Diaz2016PhotonSectorEAS,Klinkhamer2017PhotonSectorEASBound,Duenkel2021PhotonDecayEAS,Duenkel2023VacuumCherenkovEAS,Satunin2024CubicLIVShowers}.

Here we develop a simulation-based toy analysis of this effect using the reconstructed shower-maximum distributions measured by the Pierre Auger Observatory (Auger) fluorescence detector~\cite{Auger2026XmaxFD,Auger2026XmaxDataset}. We perform hybrid Monte Carlo (MC) simulations of EASs for different values of the LIV scale, fit the simulated longitudinal profiles to extract the shower maximum and calorimetric energy, and build analytical descriptions of the reconstructed observables. These distributions are then forward folded with the Auger detector acceptance and resolution parametrizations and compared with the Auger binned reconstructed shower-maximum distributions, \(X_{\max,\rm reco}\). The analysis is intended as an approximate estimate based on published \(X_{\max,\rm reco}\) distributions, rather than as a full detector-level experimental limit.

The rest of this paper is organized as follows. Section~\ref{sec:data} summarizes the fluorescence method, relevant features of the Auger experiment, and the data used in this work. Section~\ref{sec:liv} introduces the photon-sector LIV scenario and describes its expected impact on EAS development. Section~\ref{sec:simulations} describes the EAS simulations and the reconstruction of the observables used below. Section~\ref{sec:model} presents the analytical model of the reconstructed energy and \(X_{\max}\) distributions calibrated with these simulations. Section~\ref{sec:forward-folding} explains how the model is forward folded with the composition mixture and detector response to predict the Auger \(X_{\max,\rm reco}\) distributions. Section~\ref{sec:statistics} defines the likelihood-ratio test and its mock-data-set calibration. The results, including the constraint on the LIV scale and the limitations of the toy-model treatment, are discussed in Sec.~\ref{sec:results}. Section~\ref{sec:conclusions} summarizes the main conclusions.

\section{Data on the depth of air-shower maximum}
\label{sec:data}

Fluorescence telescopes detect the ultraviolet fluorescence emitted by atmospheric nitrogen molecules excited by charged shower particles (predominantly electrons and positrons). The fluorescence light intensity is approximately proportional to the deposited energy, so the atmosphere acts as a calorimeter. After corrections for light propagation, atmospheric transmission, detector response, and fluorescence yield, the measured light profile is converted into a longitudinal profile of deposited energy as a function of atmospheric depth.

This profile provides two observables relevant for the present analysis. Firstly, from the light profile, the position of the maximum energy deposit, \(X_{\max,\rm reco}\), is deduced. Secondly, the integral of the profile gives the calorimetric energy, which is then converted to the reconstructed primary energy after accounting for the invisible-energy contribution carried by particles that do not deposit their energy electromagnetically in the atmosphere. The fluorescence method therefore provides a nearly calorimetric energy estimate and direct access to the shower maximum through \(X_{\max,\rm reco}\). However, the telescopes operate on clear moonless nights only, which results in a limited duty cycle and hence in limited event statistics. In addition, quantitative reconstruction of EAS parameters is very sensitive to atmospheric conditions and relies on atmospheric monitoring, absolute calibration of the fluorescence yield, and profile reconstruction algorithms.

The present analysis uses the \(X_{\max,\rm reco}\) distributions measured with the Auger fluorescence detector and published in Ref.~\cite{Auger2026XmaxFD}. The data are provided as binned \(X_{\max,\rm reco}\) counts in intervals of reconstructed energy above \(10^{17.7}\,\text{eV}\), together with the detector-response information needed to compare model predictions with the measured distributions~\cite{Auger2026XmaxDataset}. In particular, we use the Auger \(X_{\max}\) acceptance and resolution parametrizations and the reconstructed-energy distributions within each reconstructed-energy bin. These inputs enter the forward-folding procedure described in Sec.~\ref{sec:forward-folding}. No event-level Auger data are used in this work.

\section{Lorentz invariance violation}
\label{sec:liv}

\subsection{Photon-sector Lorentz invariance violation}
\label{sec:liv:photon}

A broad class of LIV effects can be described within the Effective Field Theory (EFT) framework~\cite{Colladay:1998fq}. In this article, we concentrate on the Quantum Electrodynamics (QED) sector and consider operators in the EFT Lagrangian with mass dimensions up to six, following Refs.~\cite{Rubtsov2012LIVCrossSections, Mattingly2008LIVReview}. Additional conditions imposed on the EFT Lagrangian are discussed in Ref.~\cite{Rubtsov2012LIVCrossSections}. The EFT Lagrangian reads
\begin{equation}
    \mathscr{L} = \mathscr{L}_{\rm QED} + \mathscr{L}_e + \mathscr{L}_\gamma,
    \label{eq:eft-lagrangian}
\end{equation}
where
\(\mathscr{L}_{\rm QED}\) is the conventional QED Lagrangian.
The LIV operators associated with photons and electrons are, respectively,
\begin{align}
  & \mathscr{L}_\gamma = \frac{s_2}{4 M_{\text{LIV}}^2} F_{ij} \partial^2 F^{ij},
  \label{eq:photon-part} \\
  & \mathscr{L}_e = i \varkappa \bar{\psi} \gamma^i \mathscr{D}_i \psi + \frac{i s_{2,e}}{M_{{\rm LIV},e}^2} \mathscr{D}_j \bar{\psi} \gamma^i \mathscr{D}_i \mathscr{D}_j \psi.
  \label{eq:fermion-part}
\end{align}
Here, the covariant derivative is defined as \(\mathscr{D}_\mu \psi = \left(\partial_\mu + ie A_\mu\right) \psi\), where \(A_\mu\) denotes the electromagnetic field, \(F_{\mu\nu}\) its associated field-strength tensor, and \(\psi\) the electron field. The parameters \(M_{\rm LIV}\) and \(M_{{\rm LIV},e}\) denote the characteristic LIV mass scales in the photon and electron sectors, respectively. In general, these two scales are distinct. \(s_2\) and \(s_{2,e}\) take values \(\pm 1\), with \(+1\) corresponding to the superluminal LIV scenario for a given particle and \(-1\) to the subluminal one, while \(\varkappa\) is a dimensionless coupling parameter. Note that dimension-six operators can be induced by loop corrections: a nonzero value of \(\varkappa\) generates the subluminal correction~\eqref{eq:photon-part}; see Ref.~\cite{Satunin2017PhotonVelocityQED}.

A direct consequence of the LIV terms \eqref{eq:photon-part} and \eqref{eq:fermion-part} is the modification of the free-particle dispersion relations. In this work, we consider subluminal LIV in the photon sector and neglect LIV in the electron sector. We expect the dominant effect on EAS development to come from the suppression of photon conversion into pairs, since this process supplies most of the energetic electrons and positrons in the electromagnetic cascade. Appendix~\ref{app:elm-cascade-development} discusses the contributions of the two sectors. The corresponding photon dispersion relation is
\begin{equation}
    E_\gamma^2
    =
    k_\gamma^2
    -
    \frac{k_\gamma^4}{M_{\rm LIV}^2}.
    \label{eq:photon-liv-dispersion}
\end{equation}
Here, \(E_\gamma\) and \(k_\gamma\) denote the photon energy and the magnitude of its momentum, respectively. The complete set of modified dispersion relations is given in Ref.~\cite{Rubtsov2012LIVCrossSections}.

LIV modifies the rates of several processes relative to the LI case. For high-energy photons in the subluminal photon sector, the most relevant ones are electron--positron pair production on background photons~\cite{Jacobson2002ThresholdEffects} and in the Coulomb field of a nucleus (the Bethe--Heitler process)~\cite{Vankov2002QEDFormationLength,Rubtsov2012LIVCrossSections}. High-energy electrons (and positrons; positrons will be omitted in the following discussion for brevity), in turn, can emit photons through vacuum Cherenkov radiation, either in the soft regime (soft Cherenkov radiation) or in the hard regime (hard Cherenkov radiation), and undergo modified synchrotron radiation~\cite{Jacobson2002ThresholdEffects,Konopka2002QuantumGeometryLimits,Jacobson2006HighEnergyLIVReview}.

For subluminal LIV, the Bethe--Heitler cross section is suppressed relative to the LI case~\cite{Vankov2002QEDFormationLength,Rubtsov2012LIVCrossSections}. In the high-energy limit
\begin{equation}
E_\gamma^4/\left(2m_e^2M_{\rm LIV}^2\right)\gg1,
\label{eq:high-energy-regime-for-LIV}
\end{equation}
where \(m_e\) is the electron mass, the ratio of the LIV and LI cross sections is given by~\cite{Rubtsov2012LIVCrossSections}
\begin{equation}
    \label{eq:LIV-suppression}
    \frac{\sigma_{\rm BH,\,LIV}}{\sigma_{\rm BH,\,LI}}
    \simeq
    \frac{12m_e^2M_{\rm LIV}^2}{7E_\gamma^4}
    \ln\left(\frac{E_\gamma^4}{2m_e^2M_{\rm LIV}^2}\right).
\end{equation}
At fixed \(E_\gamma\), increasing \(M_{\rm LIV}\) takes the cross-section ratio outside the regime of validity of Eq.~\eqref{eq:LIV-suppression}. In the simulations, we use a smooth phenomenological interpolation between the standard LI cross section at weak suppression and the high-energy LIV asymptote~\cite{Martynenko2024MuonContentLIV}. This prescription recovers the LI cross section as \(M_{\rm LIV}\to+\infty\) at fixed \(E_\gamma\), so we denote the LI case by \(M_{\rm LIV}=+\infty\). To our knowledge, no first-principles calculation of the Bethe--Heitler cross section across this transition is available for the scenario considered here. The choice of interpolation is therefore a source of systematic uncertainty in the numerical bound, and we interpret the toy-model result as an order-of-magnitude estimate of the sensitivity.

The main consequence of this suppression is that photon-induced showers develop deeper in the atmosphere~\cite{Rubtsov2017TeVPhotonShowers}. Its implications for photon subshowers in hadronic showers are discussed in Sec.~\ref{sec:liv:showers}. The corresponding constraint obtained from an analysis of the shower muon content at the \(95\%\) confidence level (CL) is~\cite{Martynenko2024MuonContentLIV}
\begin{equation}
    M_{\rm LIV}>2.4\times10^{14}\,\text{GeV}.
    \label{eq:constraint-muons}
\end{equation}

Assuming an LI electron sector, hard vacuum Cherenkov emission becomes kinematically allowed above the threshold~\cite{Jacobson2002ThresholdEffects,Konopka2002QuantumGeometryLimits,Jacobson2006HighEnergyLIVReview} 
\begin{equation}
    E_e>\left(\frac{27}{4}m_e^2M_{\rm LIV}^2\right)^{1/4}.
    \label{eq:electron-scale}
\end{equation}
Here, \(E_e\) is the electron (positron) energy. For an LI electron sector, the threshold configuration contains a hard photon~\cite{Jacobson2002ThresholdEffects}. Bethe--Heitler suppression reduces the supply of electrons above threshold from photon conversion. Photons that propagate to larger depths can still convert and produce such electrons. We neglect vacuum Cherenkov emission in the toy analysis, where we expect the suppression of pair production to have the leading effect on the longitudinal profile.

We use the standard bremsstrahlung rate in the simulations. We expect small corrections to bremsstrahlung in the weak-suppression regime. In the transition regime, we expect the effect of modified bremsstrahlung to be limited because the electrons and positrons affected in this regime are produced by photons in the narrow energy interval set by the \(\propto E_\gamma^{-4}\) dependence of pair-production suppression~\eqref{eq:LIV-suppression}. In the strongly suppressed regime~\eqref{eq:high-energy-regime-for-LIV}, the effect is limited by the small fraction of photons that convert early enough for the pair to develop a cascade.

Appendix~\ref{app:elm-cascade-development} explains the motivation for these approximations and their limitations in the transition region.

The non-observation of anomalous synchrotron emission and vacuum Cherenkov radiation by electrons in the Crab Nebula constrains the electron-sector LIV scale to
\begin{equation}
M_{\rm LIV,e}>2\times10^{16}\,\text{GeV}
\label{eq:crab-constraint}
\end{equation}
at the \(95\%\) CL~\cite{Liberati2012CrabSynchrotron}. In general, an analysis sensitive to photon-sector LIV scales approaching this bound would require taking into account the full effect of electron-sector LIV in the Bethe--Heitler process~\cite{Rubtsov2012LIVCrossSections}. However, throughout this work we restrict our attention to the photon sector, assuming that the electron-sector LIV scale is sufficiently larger than the photon-sector one.

Different LIV scenarios have been investigated in earlier studies of EAS simulations. In Ref.~\cite{Antonov2001LongitudinalDevelopmentLIV}, MC simulations of hadronic EASs were carried out under the assumption of LIV leading to stable neutral pions, which in turn reduced the depth of the shower maximum. The authors of Refs.~\cite{Diaz2016PhotonSectorEAS,Klinkhamer2017PhotonSectorEASBound,Duenkel2021PhotonDecayEAS} performed MC simulations of hadronic EASs in superluminal LIV scenarios, where photon decay is kinematically allowed, and used the results to derive constraints. The impact of modified pion decay~\cite{Klinkhamer2016NeutralPionDecay} and the vacuum Cherenkov process~\cite{Duenkel2023VacuumCherenkovEAS} on EAS development has also been investigated.

In the following, we constrain \(M_{\rm LIV}\) by comparing the LIV-modified \(X_{\max,\rm reco}\) distributions with those measured by Auger.

\subsection{Expected Lorentz invariance violation effect on~hadronic shower development}
\label{sec:liv:showers}

The dominant consequence of the subluminal photon dispersion relation in Eq.~\eqref{eq:photon-liv-dispersion} is the suppression of the Bethe--Heitler pair production by high-energy photons. In an electromagnetic cascade, this delays the conversion of energetic photons into electron--positron pairs. As a result, a larger fraction of the electromagnetic energy can be transported deeper into the atmosphere before being redistributed among lower-energy particles.

This mechanism affects both the longitudinal energy-deposit profile and the energy inferred from it. For a fixed primary energy, the calorimetric energy obtained from the profile is expected to be reduced relative to the LI case, because a part of the electromagnetic energy is transported by high-energy photons to larger atmospheric depths and is not deposited in the same way as in an ordinary cascade.

The effect on \(X_{\max}\) is less direct. The delayed conversion of high-energy photons tends to move electromagnetic energy deposition deeper into the atmosphere. At the same time, the part of the electromagnetic cascade that develops after the delayed conversion contains only a fraction of the electromagnetic energy, as the highest-energy photons remain unconverted into electron--positron pairs. It can therefore reach its maximum over a shorter subsequent depth interval. At still larger depths, photons that propagated through the atmosphere with suppressed interactions can initiate secondary cascades via photoproduction, usually contributing as an extended tail of the deposited-energy profile rather than forming a dominant second maximum.

The size of these effects depends on the energies of photons in the electromagnetic cascade. Within the superposition model~\cite{Gorjunov1962EASFluctuations}, the characteristic energies of cascade particles are expected to be determined primarily by the primary energy per nucleon for nuclei-induced EASs. The energies of the photons responsible for the LIV effects should therefore follow the same scaling. Consequently, both the shift of \(X_{\max}\) and the reduction of the calorimetric energy depend primarily on the primary energy per nucleon. The LIV modification is therefore stronger for larger primary energy, smaller nuclear mass, and smaller \(M_{\rm LIV}\). This scaling motivates the use of a multidimensional simulation grid in primary energy, mass, and \(M_{\rm LIV}\), as described in Sec.~\ref{sec:simulations:setup}, and of the combined energy--mass--LIV variables introduced in Sec.~\ref{sec:model:variables}.

In the present work, these effects are propagated through the simulation-based model described in Secs.~\ref{sec:simulations} and~\ref{sec:model}. The resulting predicted distributions include both the LIV-induced change in the shower-maximum distribution and the associated bias in the reconstructed energy.

\section{Monte Carlo simulations}
\label{sec:simulations}

The analytical model used in this work is calibrated with simulations of EASs. The simulations are not intended to reproduce the full Auger fluorescence-detector response. Instead, they determine how the longitudinal EAS observables change when the photon-sector LIV modification is introduced. The detector acceptance, resolution, and binning enter only at the forward-folding stage, described in Sec.~\ref{sec:forward-folding}.

\subsection{Simulation setup}
\label{sec:simulations:setup}

We simulate nucleus-induced EASs with CONEX version 7.80~\cite{Pierog2004CONEX,Bergmann2006CONEX}, a hybrid EAS simulation code in which the first high-energy interactions are treated with an MC procedure, while the subsequent cascade evolution is described by numerical solutions of cascade equations (CE). The transition from the MC treatment to the CE description is performed at the default CONEX value \(E_{\rm MC\to CE}/E=5.0\times10^{-3}\), where \(E\) is the true primary energy. High-energy hadronic interactions are modeled with EPOS~LHC\mbox{--}R~\cite{Pierog2025EPOSLHCR}, and low-energy hadronic interactions with UrQMD~1.3~\cite{Bass1998UrQMD,Bleicher1999UrQMD}. For all CONEX simulations, the longitudinal shower profiles are forced to extend to at least \(2000\,\text{g cm}^{-2}\) in slant depth. All other CONEX configuration settings are kept at their default values.

In the present work, CONEX is used to calibrate the dependence of longitudinal EAS observables on the primary energy, nuclear mass, and LIV scale.

The simulations are performed for vertical EASs. This choice is sufficient for the present toy analysis, since the EAS development is treated as a function of slant depth \(X\) and no detailed detector geometry is modeled.

The simulated primary energies cover the interval from \(E=10^{15}\,\text{eV}\) to \(E=10^{19}\,\text{eV}\), in steps of \(\Delta\log_{10}(E/\text{eV})=0.5\). The same energy grid is used for all LIV scales and all primary nuclei.

The LIV scale is varied from \(M_{\rm LIV}=10^{11}\,\text{GeV}\) to \(10^{24}\,\text{GeV}\), in steps of \(\Delta\log_{10}(M_{\rm LIV}/\text{GeV})=1.0\). The LIV modification follows the implementation described in Ref.~\cite{Martynenko2024MuonContentLIV}. Although that implementation was originally presented for CORSIKA~\cite{Heck1998CORSIKA,Heck1998CORSIKAUserGuide} simulations, it modifies only the relevant tabulated electromagnetic interaction rates in Electron Gamma Shower version~4 (EGS4)~\cite{Nelson1988EGS4}. Since EGS4 is used to describe electromagnetic cascades in both CORSIKA and CONEX, the same modification can be applied within the CONEX hybrid framework used here. The LI reference sample is simulated separately with the unmodified electromagnetic interaction tables. In the analytical model, \(M_{\rm LIV}=+\infty\) is shorthand for this unmodified case.

The showers are simulated with \({}^{1}{\rm H}\), \({}^{4}{\rm He}\), \({}^{14}{\rm N}\), \({}^{28}{\rm Si}\), and \({}^{56}{\rm Fe}\) primary particles. This set is chosen to cover the relevant interval of nuclear masses with a reasonably uniform spacing in \(\ln(A)\), where \(A\) denotes the nuclear mass number, while keeping neighboring \(X_{\max}\) distributions sufficiently separated to remain experimentally distinguishable. The \({}^{28}{\rm Si}\) sample is used to constrain the mass-dependence of the analytical parametrization, but is not included as an independent component in the final composition fit.

For each point of the simulation grid, \(1024\) EASs are generated.

The longitudinal profiles of deposited energy \(\dd E_{\rm dep}/\dd X\) are recorded using the default CONEX slant-depth output grid retained in this work: \(200\) bin-centered values covering the interval \(0\le X\le 2000\,\text{g}\,\text{cm}^{-2}\), with a spacing of \(10\,\text{g}\,\text{cm}^{-2}\). This sampling is sufficiently fine for the profile fits used below and avoids an unnecessarily dense representation of the longitudinal profile. These profiles provide the input for the reconstruction of \(X_{\max}\) and calorimetric energy described in Sec.~\ref{sec:simulations:reconstruction}.

The simulated parameter grid is chosen to cover a broad range of the quasi-universal parameter \(\xi\equiv A^{-1}(m_e M_{\rm LIV})^{-1/2}E\)~\cite{Martynenko2024MuonContentLIV}. This parameter controls the size of the LIV effect in the parametrizations introduced in Sec.~\ref{sec:model}.

The simulation set includes primary energies well below those typically observed with the Auger fluorescence detector. These lower-energy EASs are not used to extend the data comparison in the present work, but are included to stabilize the simulation-based parametrization over a wider range of \(\xi\), which may be useful for other reconstruction approaches in future studies.

\subsection{Reconstruction of simulated profiles}
\label{sec:simulations:reconstruction}

For each simulated shower, the longitudinal profile of deposited energy is fitted with a Gaisser--Hillas function~\cite{Gaisser1977ConstantIntensityCuts} in the updated parametrization used in the Auger fluorescence-detector reconstruction~\cite{Andringa2011LongitudinalProfiles,Bellido2023LongitudinalProfiles}. The Auger normalization is expressed through the peak energy deposit. Here we reparametrize the same profile by expressing this normalization in terms of the calorimetric energy \(E_{\rm cal}\). The fitted parameters are therefore \(E_{\rm cal}\), \(X_{\max}\), and the two profile-shape parameters \(R\) and \(L\)~\cite{Andringa2011LongitudinalProfiles}. In this formulation, the fit returns both simulation-level observables used below, \(E_{\rm cal}\) and \(X_{\max}\).

The quantity denoted by \(X_{\max}\) in this section is the maximum of the fitted Gaisser--Hillas profile. It should be distinguished from \(X_{\max,\rm reco}\), which denotes the detector-level reconstructed observable introduced later in the forward-folding procedure. We do not use the maximum of the discrete CONEX profile to determine \(X_{\max}\), since fitting the profile is methodologically closer to the reconstruction procedure used in the Auger fluorescence-detector analyses.

The Gaisser--Hillas fit is performed with regularization terms for the shape parameters \(R\) and \(L\), following the constrained-profile strategy used in Auger fluorescence-detector reconstructions~\cite{Bellido2023LongitudinalProfiles}. These regularization terms stabilize the profile fit while still leaving the profile shape free to respond to the LIV-induced modifications. The explicit form of the fitted profile, likelihood, and regularization terms, together with the details of the fitting procedure, are given in Appendix~\ref{app:profile-reconstruction}.

To avoid partial double counting of detector-response effects, all recorded CONEX depth points are retained in the profile fit, and no additional quality cuts or fit-range restrictions are imposed at this stage. The only selection applied to the simulation-level reconstruction is the removal of fits for which the numerical minimization of the corresponding negative log-likelihood did not converge. Overall, \(170\,919\) of the \(691\,200\) simulated profiles (\(24.7\%\)) are excluded because the fit fails to converge. Auger reports a profile reconstruction failure rate of \(16.9\%\)~\cite{Auger2026XmaxFD}. The comparison is only approximate because the Auger rate reflects the unknown true cosmic-ray spectrum and composition together with observatory's event pre-selection, and our rate is computed over an \(M_{\rm LIV}\) grid. The fraction of converged fits per simulation grid point (\(1024\) showers) ranges from \(72\%\) to \(82\%\) for \(\xi<10\) and falls to \(60\%\)--\(66\%\) in the strongest simulated LIV regime (\(\xi\sim10^6\)). The experimental acceptance, resolution, and Auger selection effects are included later in the forward-folding procedure described in Sec.~\ref{sec:forward-folding}.

\section{Analytical model of reconstructed observables}
\label{sec:model}

\subsection{Variables}
\label{sec:model:variables}

For the analytical parametrization, it is convenient to use logarithmic energy variables normalized to \(10^{19}\,\text{eV}\). We define
\begin{equation}
    \begin{aligned}
    \epsilon_{\rm cal}
    &\equiv
    \log_{10}\left(\frac{E_{\rm cal}}{10^{19}\,\text{eV}}\right),
    \\
    \epsilon
    &\equiv
    \log_{10}\left(\frac{E}{10^{19}\,\text{eV}}\right),
    \\
    \epsilon_{\rm reco}
    &\equiv
    \log_{10}\left(\frac{E_{\rm reco}}{10^{19}\,\text{eV}}\right),
    \end{aligned}
    \label{eq:energy-variables}
\end{equation}
where \(E\) is the true primary energy defined in Sec.~\ref{sec:simulations:setup}, \(E_{\rm cal}\) is the calorimetric energy obtained from the Gaisser--Hillas fit in Sec.~\ref{sec:simulations:reconstruction}, and \(E_{\rm reco}\) is the energy inferred from \(E_{\rm cal}\) using the LI energy-reconstruction relation introduced below. The primary mass is represented by
\begin{equation}
    Y\equiv\ln(A).
    \label{eq:mass-variable}
\end{equation}

The LIV scale is expressed through
\begin{equation}
    \eta
    \equiv
    \log_{10}
    \left[
        \frac{(m_e M_{\rm LIV})^{1/2}}{10^{19}\,\text{eV}}
    \right].
    \label{eq:eta-definition}
\end{equation}
With this convention, smaller values of \(\eta\) correspond to stronger LIV effects, while the LI limit is recovered for \(\eta\to+\infty\).

The dependence on primary energy, mass, and LIV scale is described by the dimensionless parameter \(\xi\) introduced in Sec.~\ref{sec:simulations:setup}. In logarithmic form,
\begin{equation}
    \zeta
    \equiv
    \log_{10}\xi
    =
    \epsilon-\eta-\frac{Y}{\ln 10}.
    \label{eq:zeta-definition}
\end{equation}

For quantities expressed in terms of the reconstructed energy, we use analogous variables,
\begin{equation}
    \xi_{\rm reco}
    \equiv
    (m_e M_{\rm LIV})^{-1/2}A^{-1}E_{\rm reco},
    \label{eq:xi-reco-definition}
\end{equation}
and
\begin{equation}
    \zeta_{\rm reco}
    \equiv
    \log_{10}\xi_{\rm reco}
    =
    \epsilon_{\rm reco}-\eta-\frac{Y}{\ln 10}.
    \label{eq:zeta-reco-definition}
\end{equation}

These variables are used below to describe the reconstructed-energy bias and the \(X_{\max}\) distribution in a form common to different primary masses and LIV scales.

\subsection{Reconstructed-energy bias}
\label{sec:model:energy}

The reconstructed energy used in the forward model is defined by applying an LI calibration to the calorimetric energy obtained from the fitted longitudinal profile. In logarithmic variables, this relation is written as
\begin{equation}
    \epsilon_{\rm reco}
    =
    \epsilon_{\rm offset}
    +
    \kappa_{\rm cal}\,\epsilon_{\rm cal},
    \label{eq:energy-reco-li}
\end{equation}
where \(\epsilon_{\rm offset}\) is the logarithmic energy offset and \(\kappa_{\rm cal}\) is the calorimetric-energy calibration slope. The values fitted to the LI simulations are
\begin{equation}
    \epsilon_{\rm offset}=3.95\times10^{-2},
    \qquad
    \kappa_{\rm cal}=9.79\times10^{-1}.
    \label{eq:energy-reco-li-parameters}
\end{equation}
By construction, Eq.~\eqref{eq:energy-reco-li} gives an approximately unbiased estimate of the true primary energy in the LI case, up to the residual scatter of the simulation-level reconstruction.

In the LIV case, the same LI calibration is applied to the calorimetric energy. This prescription mimics applying a standard fluorescence-detector energy-reconstruction procedure to EASs whose electromagnetic component is modified by LIV. The LIV-induced modification of the electromagnetic cascade reduces the calorimetric energy at fixed primary energy, thereby introducing a systematic downward bias in the reconstructed energy. We parametrize this effect as a function of \(\zeta_{\rm reco}\), rather than \(\zeta\), because the comparison with data is performed in bins of reconstructed energy,
\begin{equation}
    \epsilon-\epsilon_{\rm reco}
    =
    \Delta_{\epsilon}\,
    \ln\left[
        1+
        \exp\left(
            \frac{\zeta_{\rm reco}-\zeta_{\epsilon,c}}
                 {\zeta_{\epsilon,s}}
        \right)
    \right].
    \label{eq:energy-bias}
\end{equation}
Here, \(\Delta_\epsilon\), \(\zeta_{\epsilon,c}\), and \(\zeta_{\epsilon,s}\) are the amplitude, onset, and transition-width parameters of the reconstructed-energy bias, respectively.

\begin{figure}
    \centering
    \includegraphics[width=\linewidth]{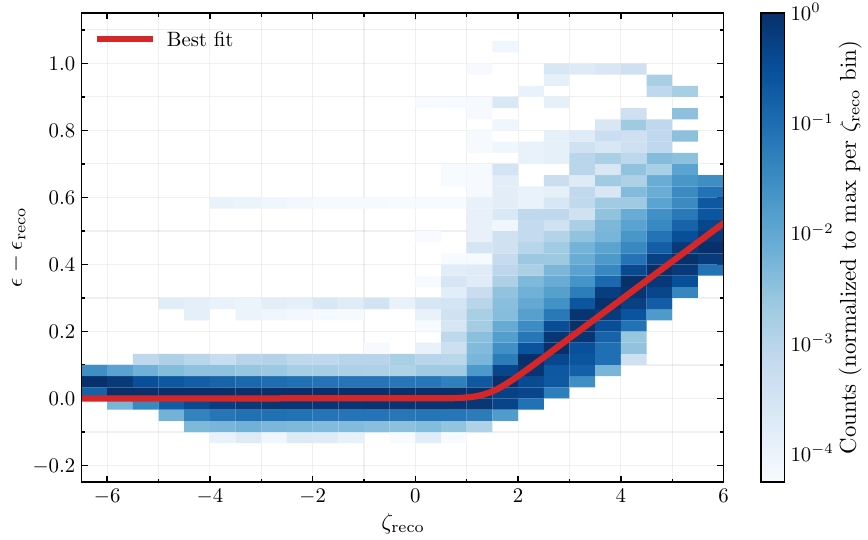}
    \caption{Reconstructed-energy bias in the LIV simulations as a function of \(\zeta_{\rm reco}\). The blue color scale shows a column-normalized two-dimensional histogram of simulated EASs in the \((\zeta_{\rm reco},\,\epsilon-\epsilon_{\rm reco})\) plane. In each \(\zeta_{\rm reco}\) bin, the bin counts are normalized to the maximum count in the corresponding vertical slice. The solid red curve shows Eq.~\eqref{eq:energy-bias} with the best-fit parameters.}
    \label{fig:energy-bias}
\end{figure}

The heuristic parametrization in Eq.~\eqref{eq:energy-bias} reflects the trend seen in simulations: the energy bias grows linearly with \(\zeta_{\rm reco}\) in the strong-LIV limit and vanishes in the LI limit. The fitted parameters are 
\begin{equation}
    \Delta_{\epsilon}=2.41\times 10^{-2},
    \qquad
    \zeta_{\epsilon,c}=1.43,
    \qquad
    \zeta_{\epsilon,s}=2.11\times 10^{-1}.
    \label{eq:energy-bias-parameters}
\end{equation}
Figure~\ref{fig:energy-bias} shows that this parametrization captures the onset and subsequent growth of the reconstructed-energy bias over the simulated range of \(\zeta_{\rm reco}\).

Equations~\eqref{eq:energy-reco-li} and~\eqref{eq:energy-bias} define the mapping between true and reconstructed logarithmic energies used in the analytical model. For a given reconstructed energy, primary mass, and LIV scale, the corresponding true energy is obtained as
\begin{equation}
    \epsilon
    =
    \epsilon_{\rm reco}
    +
    \Delta_{\epsilon}\,
    \ln\left[
        1+
        \exp\left(
            \frac{
                \epsilon_{\rm reco}
                -
                \eta
                -
                Y/\ln 10
                -
                \zeta_{\epsilon,c}
            }
            {\zeta_{\epsilon,s}}
        \right)
    \right].
    \label{eq:true-energy-from-reco}
\end{equation}
This form is used below when the \(X_{\max}\) distribution is evaluated for a fixed reconstructed-energy bin.

The scatter of the reconstructed energy around the parametrized relation is treated as an additional response term. The residual scatter obtained from the simulation-based fit is evaluated in \(\epsilon_{\rm reco}\). For the forward-folding step, this scatter is converted to an approximate relative energy scatter by multiplying it by \(\ln (10)\), which is sufficient for the level of accuracy of the present analysis. This gives a residual relative energy resolution of \(7.5\%\). This contribution is added in quadrature to the \(14\%\) systematic uncertainty in the energy scale reported by Auger for the \(X_{\max,\rm reco}\) distributions. The combined reconstructed-energy resolution is not used to smear the shower-maximum distributions, because the comparison is performed with the Auger reconstructed-energy distributions. Instead, it is used when assigning probability uncertainties for the mock-data-set calibration. The covariance matrix of the fitted parameters is also used only later in the mock-data-set calibration to account for the statistical uncertainty of the analytical parametrization; see Sec.~\ref{sec:statistics:mock-data-sets}.

\subsection{Shower-maximum distribution}
\label{sec:model:xmax}

The distribution of \(X_{\max}\) at fixed primary energy and mass is described with a generalized Gumbel distribution, which naturally arises from the Heitler--Matthews model~\cite{Matthews2005HeitlerModel} of EAS longitudinal development, as shown in Ref.~\cite{DeDomenico2013EASDevelopment}. In this work, we follow the parametrization used in Refs.~\cite{DeDomenico2013EASDevelopment,Evoli2026XmaxMoments}. We apply this model to the simulation-level \(X_{\max}\) values reconstructed from the fitted Gaisser--Hillas profiles described in Sec.~\ref{sec:simulations:reconstruction}. For a primary with logarithmic energy \(\epsilon\), logarithmic mass \(Y=\ln(A)\), and LIV parameter \(\eta\), the corresponding probability density is denoted by
\begin{equation}
    {\cal G}(X_{\max}\mid\epsilon,Y,\eta).
    \label{eq:xmax-gumbel-density-symbol}
\end{equation}
The explicit generalized-Gumbel form and the LI-limit energy--mass parametrization are given in Appendix~\ref{app:xmax-parametrization}.

In the LIV case, the CONEX simulations show that, within the accuracy needed for the present model, the dominant modification of the \(X_{\max}\) distribution can be described by a shift of the location parameter \(\mu\). We therefore keep the generalized-Gumbel shape and width parameters equal to their LI-limit forms at the same \((\epsilon,Y)\), and introduce LIV only through the shift \begin{equation}
    \mu(\epsilon,Y,\eta)
    =
    \mu_{\rm LI}(\epsilon,Y)
    +
    \delta\mu_{\rm LIV}(\zeta,\eta).
    \label{eq:xmax-mu-liv}
\end{equation}
The LIV-induced shift is parametrized as
\begin{equation}
    \begin{aligned}
    \delta\mu_{\rm LIV}(\zeta,\eta)
    &=
    \operatorname{expit}\left(\frac{\zeta-\zeta_{\mu,c}}{\zeta_{\mu,s}}
    \right)\times\\
    &\times
    \left[
        \exp\left(-\frac{\eta}{\eta_\mu}\right)
        \Delta_\mu^+
        -
        \left(\zeta-\zeta_{\mu,c}\right)
        \Delta_\mu^-
    \right],
    \end{aligned}
    \label{eq:xmax-mu-shift}
\end{equation} where \(\operatorname{expit}(z) \equiv \left[1+\exp(-z)\right]^{-1}\) provides a smooth interpolation between the LI and strong-LIV limiting regimes. The positive term in square brackets describes the deepening of the electromagnetic-cascade starting point due to LIV, while the negative term reflects the fact that the LIV electromagnetic cascade needs a shorter depth interval before reaching its maximum. The variable \(\zeta\) captures the leading quasi-universal scaling of the suppression of pair production. The cascade development, however, also depends on the absolute particle energies and is therefore not fully determined by \(\zeta\). Consistently, the simulations show a residual dependence of the \(X_{\max}\) shift on \(\eta\) at fixed \(\zeta\), which is included empirically in Eq.~\eqref{eq:xmax-mu-shift}. In total, there are five fitted parameters controlling the LIV-induced shift of the Gumbel location parameter: amplitudes \(\Delta_{\mu}^{+}\) and \(\Delta_{\mu}^{-}\), onset \(\zeta_{\mu,c}\), transition width \(\zeta_{\mu,s}\), and saturation scale \(\eta_{\mu}\).

Here, in contrast to~\eqref{eq:energy-bias}, \(\zeta\), rather than \(\zeta_{\rm reco}\), is used because \({\cal G}(X_{\max}\mid\epsilon,Y,\eta)\) describes the EAS distribution at a fixed true primary energy before folding with the detector response. The conversion from reconstructed energy to the corresponding true energy is applied later, when the model is evaluated in the Auger reconstructed-energy bins.

The parameters of the full \(X_{\max}\) model are obtained by minimizing the negative log-likelihood of the reconstructed \(X_{\max}\) samples from the simulations, following the strategy of Ref.~\cite{Evoli2026XmaxMoments}. In contrast to the purely LI parametrization of Ref.~\cite{Evoli2026XmaxMoments}, the fit used here includes both the LI-limit coefficient matrices and the LIV-shift parameters. Thus, all 26 parameters of the \(X_{\max}\) model are fitted simultaneously over the simulated \((E,A,M_{\rm LIV})\) grid, including the LI reference sample. The details are provided in Appendix~\ref{app:xmax-parametrization}.

The fitted LIV-shift parameters are 
\begin{equation}
    \begin{aligned}
    &\Delta_\mu^+=8.41\times 10^{-2}\,\text{g}\,\text{cm}^{-2},
    \quad
    \Delta_\mu^-=7.34\,\text{g}\,\text{cm}^{-2},
    \\
    &\eta_\mu=1.04,
    \quad
    \zeta_{\mu,c}=2.45,
    \quad
    \zeta_{\mu,s}=7.33\times 10^{-1}.
    \end{aligned}
    \label{eq:xmax-mu-shift-parameters}
\end{equation}

\begin{figure}[htb]
    \centering
    \includegraphics[width=\linewidth]{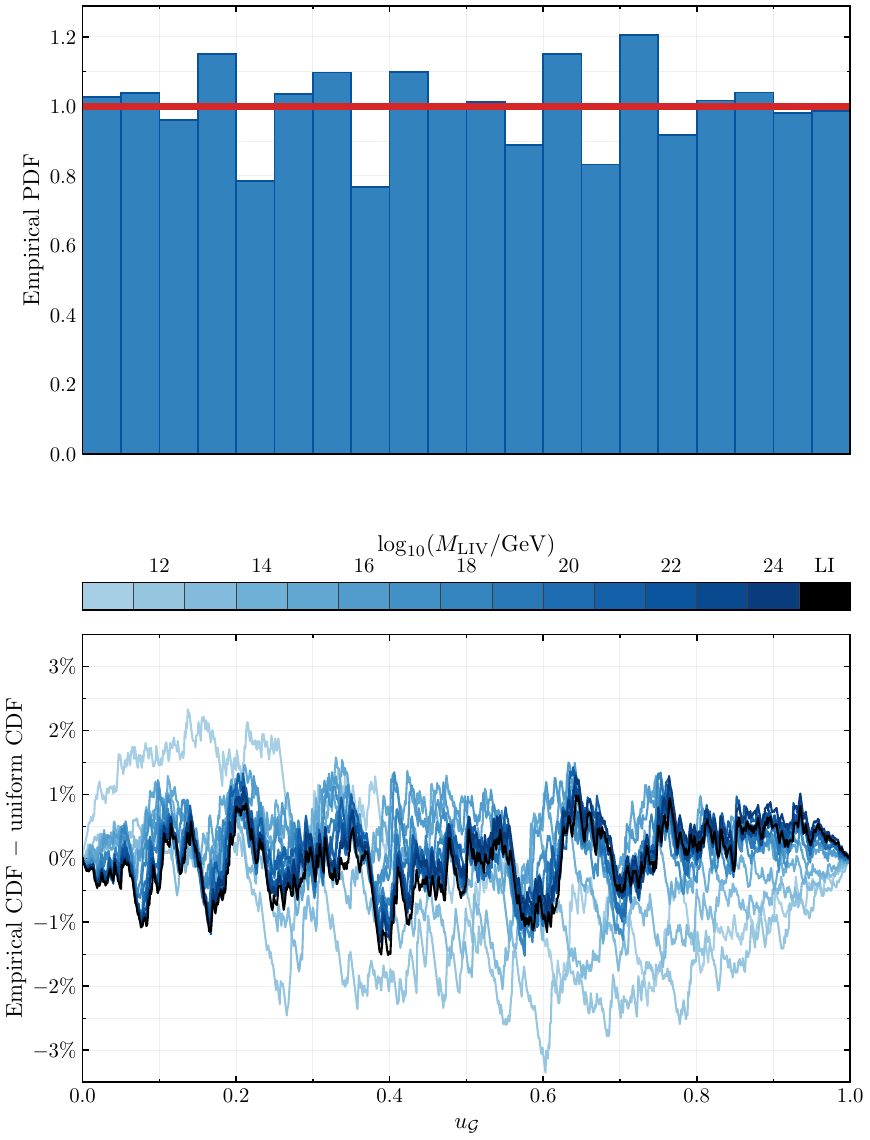}
    \caption{In-sample comparison of the analytical and simulated \(X_{\max}\) distributions. The upper panel shows the empirical probability density function (PDF) of \(u_{\mathcal G}\) for all showers used in the fit. The red line shows the uniform PDF on \([0,1]\). The lower panel shows the empirical CDF's signed departure from uniformity at each \(\eta\), combining showers across all \((\epsilon,Y)\). The color scale gives the corresponding values of \(\log_{10}(M_{\rm LIV}/\text{GeV})\). The black curve denotes the LI case.}
    \label{fig:in-sample-validation-xmax}
\end{figure}

For each simulated EAS, we evaluate the fitted generalized-Gumbel cumulative distribution function
(CDF) at the Gaisser--Hillas \(X_{\max}\),
\begin{equation}
    u_{\mathcal G} \equiv
    \int_{-\infty}^{X_{\max}}
    \dd X'_{\max}\,\mathcal G(X'_{\max}\mid\epsilon,Y,\eta),
    \label{eq:uniform-variable}
\end{equation}
using the simulation parameters \(\epsilon\), \(Y\), and \(\eta\) of that shower. If the parametrization reproduces the simulated \(X_{\max}\) distributions exactly, \(u_{\mathcal G}\) is uniform on \([0,1]\). Figure~\ref{fig:in-sample-validation-xmax} shows the departures from uniformity in the CONEX sample used for the fit.

No additional shower-level smearing of \(X_{\max}\) is introduced at this stage. Shower-to-shower fluctuations are already encoded in the generalized-Gumbel distribution fitted to the CONEX samples. The detector acceptance and resolution are applied separately in the forward-folding procedure.
The residual shape differences seen in
Fig.~\ref{fig:in-sample-validation-xmax} are further used to estimate an additional
contribution to the width of the mock likelihood-ratio distribution. A detailed description is given below.

\section{Forward folding to reconstructed shower maxima}
\label{sec:forward-folding}

\subsection{Composition mixture}
\label{sec:forward-folding:composition}

The forward-folding model converts the continuous \(X_{\max}\) distributions of Sec.~\ref{sec:model:xmax} into predicted event probabilities in the Auger bins. We use the same reconstructed-energy and \(X_{\max,\rm reco}\) binning as the Auger distributions. Hereafter, \(x\) labels the Auger \(X_{\max,\rm reco}\) bins and \(b\) labels the Auger reconstructed-energy bins.

The final composition model is built from four nuclear components corresponding to \({}^{1}{\rm H}\), \({}^{4}{\rm He}\), \({}^{14}{\rm N}\), and \({}^{56}{\rm Fe}\). Hereafter, the index \(y\in\{{}^{1}{\rm H},{}^{4}{\rm He},{}^{14}{\rm N},{}^{56}{\rm Fe}\}\) labels the four fitted mass groups.

We follow the structure of the Auger composition analysis and allow the composition fractions to vary independently in each reconstructed-energy bin. The corresponding composition vector in bin \(b\) is denoted by
\begin{equation}
    \mathbf f_b
    \equiv
    \left\{
        f_{b,y} \mid y \in 
        \{{}^{1}{\rm H},
        {}^{4}{\rm He},
        {}^{14}{\rm N},
        {}^{56}{\rm Fe}\}
    \right\}.
    \label{eq:composition-vector}
\end{equation}
Its components satisfy
\begin{equation}
    f_{b,y}\ge 0,
    \qquad
    \sum_y f_{b,y}=1
    \label{eq:composition-simplex}
\end{equation}
for every reconstructed-energy bin \(b\).

The full set of composition fractions is denoted by \(F\). It is an array formed by the composition vectors in the reconstructed-energy bins,
\begin{equation}
    F\equiv\{\mathbf f_b\}.
    \label{eq:composition-matrix}
\end{equation}
No smoothness condition or common energy dependence is imposed across reconstructed-energy bins. This choice keeps the comparison close to the Auger composition-fit procedure and isolates the effect of replacing the LI EAS predictions by the corresponding LIV predictions.

For a fixed LIV scale \(\eta\), the mixture probability \({\cal P}_{x,b}(\eta,\mathbf{f}_b)\) in \(X_{\max,\rm reco}\) bin \(x\) and reconstructed-energy bin \(b\) is written as
\begin{equation}
    {\cal P}_{x,b}(\eta,\mathbf{f}_b)
    =
    \sum_y
    f_{b,y}\,
    {\cal P}_{x,b,y}(\eta),
    \label{eq:composition-mixture}
\end{equation}
where \({\cal P}_{x,b,y}(\eta)\) is the forward-folded probability for mass group \(y\). The detector folding and bin integration used to construct these component probabilities are described in Secs.~\ref{sec:forward-folding:detector} and~\ref{sec:forward-folding:bins}.

\subsection{Detector acceptance and resolution}
\label{sec:forward-folding:detector}

The fluorescence-detector response is implemented using the acceptance and resolution parametrizations provided with the Auger reconstructed-shower-maximum distributions~\cite{Auger2026XmaxFD,Auger2026XmaxDataset}. We use these parametrizations as fixed external inputs and apply them in the same way to the LI and LIV predictions. The explicit response functions and the notation used for their tabulated parameters are summarized in Appendix~\ref{app:auger-response}.

At fixed reconstructed logarithmic energy \(\epsilon_{\rm reco}\), mass group \(y\), and LIV scale \(\eta\), the true logarithmic energy is obtained from Eq.~\eqref{eq:true-energy-from-reco},
\begin{equation}
    \epsilon=\epsilon(\epsilon_{\rm reco},Y_y,\eta).
    \label{eq:forward-true-energy}
\end{equation}
The shower-level distribution \({\cal G}(X_{\max}\mid\epsilon,Y_y,\eta)\) is then multiplied by the Auger \(X_{\max}\) acceptance and convolved with the Auger \(X_{\max}\) resolution. The acceptance is applied to the simulation-level fitted \(X_{\max}\), while the resolution maps this quantity to the detector-level observable \(X_{\max,\rm reco}\).

The resulting conditional probability density for a single mass group is
\begin{widetext}
\begin{equation}
    \frac{\dd {\cal P}_{b,y}}{\dd X_{\max,\rm reco}}(X_{\max,\rm reco}\mid\epsilon_{\rm reco},Y_y,\eta)
    =
    \int \dd X_{\max}\,
    {\cal R}_{b}(X_{\max,\rm reco}-X_{\max})
    \frac{
        {\cal A}_{b}(X_{\max})
        {\cal G}(X_{\max}\mid\epsilon(\epsilon_{\rm reco},Y_y,\eta),Y_y,\eta)
    }{
        \int \dd X_{\max}'\,
        {\cal A}_{b}(X_{\max}')
        {\cal G}(X_{\max}'\mid\epsilon(\epsilon_{\rm reco},Y_y,\eta),Y_y,\eta)
    }.
    \label{eq:forward-folded-density}
\end{equation}
\end{widetext}
Here \({\cal A}_{b}\) is the acceptance in reconstructed-energy bin \(b\), \({\cal R}_{b}\) is the corresponding shower-maximum-resolution kernel, and the denominator normalizes the accepted shower-level \(X_{\max}\) distribution before convolution with the detector resolution.

\subsection{Binned probabilities}
\label{sec:forward-folding:bins}

The final component probabilities are obtained by integrating the detector-folded density of Sec.~\ref{sec:forward-folding:detector} over the Auger reconstructed-energy and \(X_{\max,\rm reco}\) bins. For a mass group \(y\), the probability in \(X_{\max,\rm reco}\) bin \(x\) and reconstructed-energy bin \(b\) is
\begin{widetext}
\begin{equation}
    {\cal P}_{x,b,y}(\eta)
    =
    \int\limits_{b\text{ bin}}
    \dd\epsilon_{\rm reco}\,
    w_b(\epsilon_{\rm reco})
    \int\limits_{x\text{ bin}}
    \dd X_{\max,\rm reco}\,
    \frac{\dd {\cal P}_{b,y}}{\dd X_{\max,\rm reco}}
    (X_{\max,\rm reco}\mid\epsilon_{\rm reco},Y_y,\eta).
    \label{eq:binned-mass-group-probability}
\end{equation}
\end{widetext}
Here \(w_b(\epsilon_{\rm reco})\) describes the local distribution of reconstructed energies inside Auger reconstructed-energy bin \(b\) and is normalized to unity over that bin. This distribution is provided in Ref.~\cite{Auger2026XmaxFD} as a local power law after the pre-selection procedure and therefore should not be interpreted as the physical cosmic-ray spectrum.

The mixed-composition probability \({\cal P}_{x,b}(\eta,\mathbf{f}_b)\) is then obtained from Eq.~\eqref{eq:composition-mixture}. The predicted probabilities are normalized to the observed number of events in each reconstructed-energy bin. The observed number of events in \(X_{\max,\rm reco}\) bin \(x\) and reconstructed-energy bin \(b\) is denoted by \(n_{x,b}\). The total observed number of events in reconstructed-energy bin \(b\) is
\begin{equation}
    N_{b,\rm obs}=\sum_x n_{x,b}.
    \label{eq:observed-events-energy-bin}
\end{equation}
The expected number of events in the same bin is then
\begin{equation}
    \nu_{x,b}(\eta,\mathbf{f}_b)
    =
    N_{b,\rm obs}
    \frac{
        {\cal P}_{x,b}(\eta,\mathbf{f}_b)
    }{
        \sum\limits_{x'}{\cal P}_{x',b}(\eta,\mathbf{f}_b)
    }.
    \label{eq:expected-counts}
\end{equation}
This normalization removes the need to model the absolute exposure and flux normalization. The likelihood used below is therefore sensitive only to the shape of the \(X_{\max,\rm reco}\) distribution in each reconstructed-energy bin.

The integrals in Eq.~\eqref{eq:binned-mass-group-probability} are evaluated numerically for each mass group, reconstructed-energy bin, \(X_{\max,\rm reco}\) bin, and LIV scale.

In addition to the central probabilities, we estimate the corresponding uncertainties by propagating the covariance matrices of the analytical energy and \(X_{\max}\) parametrizations, obtained from the corresponding fits, through the probability model. The propagation is performed by contracting these covariance matrices with the gradients of the predicted probabilities with respect to the model parameters. We further include the uncertainties of the Auger detector-response parameters, the Auger energy-resolution contribution combined with the residual simulation-level reconstructed-energy response, and the numerical integration error. These probability uncertainties are used in the mock-data-set calibration described in Sec.~\ref{sec:statistics:mock-data-sets}.

\section{Statistical procedure}
\label{sec:statistics}

\subsection{Likelihood-ratio statistic}
\label{sec:statistics:likelihood-ratio}

The statistical comparison is based on the binned \(X_{\max,\rm reco}\) counts introduced in Sec.~\ref{sec:forward-folding:bins}. For each tested LIV scale \(\eta\), the composition fractions collected in \(F\), defined in Eq.~\eqref{eq:composition-matrix}, are fitted to the data. The array \(F\) contains \(20\times4\) fractions, constrained by positivity and by one normalization condition in each reconstructed-energy bin, as specified in Eq.~\eqref{eq:composition-simplex}. The normalization constraints leave \(60\) independent composition degrees of freedom.

For each tested \(\eta\), the composition fractions are obtained with an Auger-style binned maximum-likelihood fit of the predicted \(X_{\max,\rm reco}\) distributions to the observed bin counts. Following the Auger composition analysis~\cite{Auger2014XmaxComposition}, the fitted quantity is written as a likelihood ratio with respect to the saturated model, for which the predicted bin counts are equal to the observed ones. The explicit likelihood expression and the constrained-fit details are given in Appendix~\ref{app:statistical-procedure}.

The fitted array \(\widehat F(\eta)\) is used only to account for the unknown mass composition and is not interpreted as a physical composition measurement. One reason is that the distribution model contains only four representative mass groups, whereas the true cosmic-ray composition may contain a richer mixture of nuclei. This limitation is especially relevant under the LIV hypothesis, where the LIV-induced bias in \(X_{\max,\rm reco}\) can change the mapping between nuclear mass and the observed \(X_{\max,\rm reco}\) distribution. As a result, four LI-motivated representative groups need not provide a complete representation of the physical composition. A second reason is that no smoothness constraint is imposed between neighboring reconstructed-energy bins. This choice follows the Auger composition-fit strategy, but further limits any direct physical interpretation of the fitted fractions under the LIV hypothesis.

After fitting \(F\), the test statistic is constructed from the ordinary unweighted Poisson likelihood. The LI hypothesis is represented by the limiting case \(\eta=+\infty\). For each finite LIV scale, we compare the finite-\(\eta\) fit with the LI fit using
\begin{equation}
    D(\eta)
    \equiv
    -2\left(\ln{\cal L}_\eta-\ln{\cal L}_{+\infty}\right).
    \label{eq:lrt-statistic}
\end{equation}
Here, \(\ln{\cal L}_\eta\) and \(\ln{\cal L}_{+\infty}\) are the unweighted Poisson log-likelihoods after composition fitting for the finite-\(\eta\) and LI hypotheses, respectively. Positive values of \(D(\eta)\) indicate a preference of the observed \(X_{\max,\rm reco}\) distributions for the fitted LI prediction over the fitted finite-\(\eta\) LIV prediction. The statistical interpretation of \(D\) is obtained by the mock-data-set calibration described in Sec.~\ref{sec:statistics:mock-data-sets}, rather than by assuming an asymptotic \(\chi^2\) distribution.

\subsection{Mock-data-set calibration}
\label{sec:statistics:mock-data-sets}

The distribution of the statistic \(D(\eta)\) is calibrated with mock data sets generated separately for each tested finite value of \(\eta\). We do not use the asymptotic Wilks approximation because the test compares a finite-LIV hypothesis with the LI hypothesis, rather than nested models with different numbers of free parameters. In addition, the test includes bounded composition fractions and depends on predictions derived from finite simulation samples with non-negligible parametrization and detector-response uncertainties.

For a fixed finite value of \(\eta\), the mock data sets are generated from the fitted LIV prediction at that same value of \(\eta\). The composition fractions are set to \(\widehat F(\eta)\), while the single-mass-group probabilities are fluctuated according to their estimated uncertainties before generating the mock bin counts. The fluctuated probabilities are used only to generate the mock data sets; the subsequent fits use the nominal probabilities, as in the fit to the observed data.

Each mock data set is refitted under both the tested finite-\(\eta\) hypothesis and the LI hypothesis, using the same fitting procedure as for the observed data. This gives an empirical distribution of \(D(\eta)\) under the tested finite-LIV hypothesis. 

Residual errors in the parametrized \(X_{\max}\) distributions can redistribute events among reconstructed depth bins and thereby change the likelihood-ratio statistic. The effect depends on the tested \(\eta\), since the depth bins that distinguish the LIV prediction from the LI prediction, and their relative contributions to the likelihood ratio, vary with \(\eta\). We use the discrepancies between the fitted and simulated distributions to estimate this additional uncertainty and broaden the mock distribution of \(D(\eta)\) around its mean at each tested scale.

The calibrated \(p\)-value is the corresponding upper-tail probability for obtaining a broadened value of \(D(\eta)\) at least as large as the observed one. The broadening is conservative for an observed statistic at or above the mock mean, since it can only increase or leave unchanged the upper-tail \(p\)-value. The detailed mock-data generation procedure, including the treatment of probability uncertainties, \(D(\eta)\) broadening, and the finite-sample regularization of the \(p\)-value, is given in Appendix~\ref{app:statistical-procedure}.

The calibrated \(p\)-value has the following interpretation: under the tested finite-\(\eta\) hypothesis, it is the probability of obtaining a likelihood-ratio statistic at least as unfavorable to that hypothesis as the observed one. A finite LIV scale is excluded at CL \((1-\alpha)\) when
\begin{equation}
    p(\eta)<\alpha.
    \label{eq:exclusion-criterion}
\end{equation}
The lower bound on the LIV scale is obtained from the crossing of the calibrated \(p(\eta)\) curve with the chosen CL threshold, after converting \(\eta\) to \(M_{\rm LIV}\) using Eq.~\eqref{eq:eta-definition}.

\section{Results and discussion}
\label{sec:results}

\subsection{Constraint on the scale of Lorentz invariance violation}
\label{sec:results:constraint}

The statistical procedure of Sec.~\ref{sec:statistics} is applied to the Auger \(X_{\max,\rm reco}\) distributions using the forward-folded predictions described in Sec.~\ref{sec:forward-folding}. For each tested finite value of \(\eta\), the composition fractions in all reconstructed-energy bins are fitted simultaneously, with independent normalization constraints imposed in each bin. Although the weighted objective function separates into independent contributions from the reconstructed-energy bins for fixed \(\eta\), the composition is determined in one simultaneous fit. The resulting LIV fit is then compared with the LI limit using the statistic \(D(\eta)\).

\begin{figure*}[htb]
    \centering
    \includegraphics[height=0.3\linewidth]{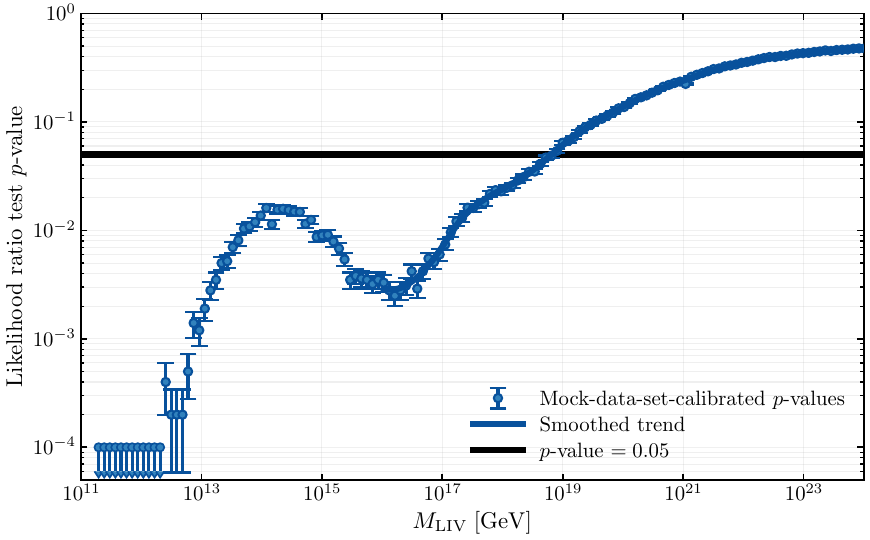}\hfill\includegraphics[height=0.3\linewidth]{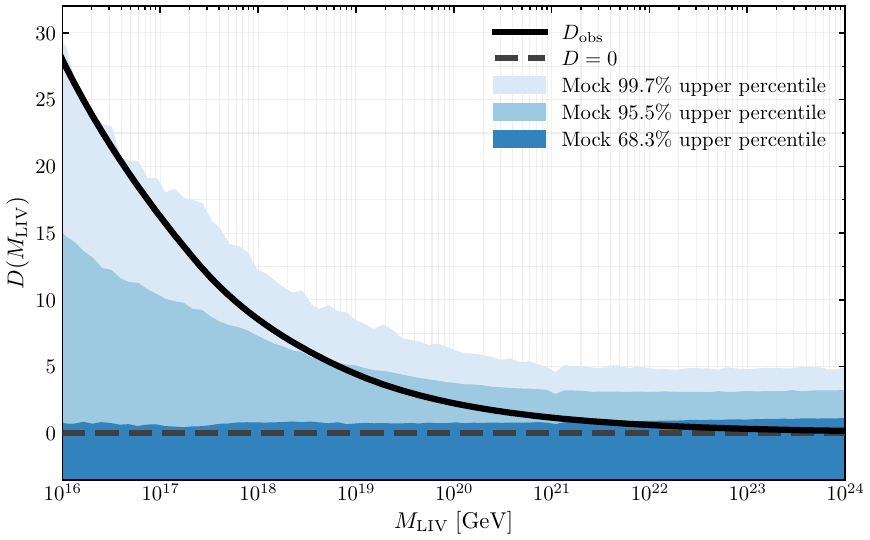}
    \caption{Likelihood-ratio constraint on the LIV scale. Left panel: Points show the \(p\)-values obtained from the mock-data-set calibration described in Sec.~\ref{sec:statistics:mock-data-sets}. The error bars show the estimated sampling uncertainty due to the finite number of mock data sets. The curve is a smoothing spline fitted to \(\ln[p(\eta)/0.05]\) as a function of \(\log_{10}(M_{\rm LIV}/\text{GeV})\) for \(M_{\rm LIV}\ge10^{16}\,\text{GeV}\) and transformed back to \(p\). Its crossing of the horizontal line at \(p=0.05\) gives the approximate \(95\%\) CL lower bound. Values of \(M_{\rm LIV}\) below the crossing are excluded within the present toy-model treatment. Points at the lowest displayed \(p\)-values indicate saturation of our regularized estimate: no mock data set among \(10^{4}\) generated under the tested finite-LIV hypothesis produced a preference for the LI limit as strong as that observed in the data, giving \(p\simeq 10^{-4}\). Right panel: Likelihood-ratio statistic as a function of the LIV scale. The black curve shows the observed statistic \(D_{\rm obs}(M_{\rm LIV})\) defined in Eq.~\eqref{eq:lrt-statistic}. The blue regions show upper percentiles of broadened \(D(M_{\rm LIV})\) obtained from mock data sets generated under the corresponding finite-LIV hypothesis. The dashed line marks \(D=0\), where the finite-LIV and LI predictions give equal Poisson likelihoods after composition fitting. Both panels use the analytical model of Sec.~\ref{sec:model}, forward folded as described in Sec.~\ref{sec:forward-folding}. The model is calibrated on the coarser grid of CONEX shower simulations described in Sec.~\ref{sec:simulations:setup}.}
    \label{fig:lrt-scan}
\end{figure*}

The scan results are shown in Fig.~\ref{fig:lrt-scan}. We evaluate the forward-folded analytical model on a grid of \(141\) equally spaced values of \(\eta\) from \(-6\) to \(0.5\). At each tested scale, we calibrate the statistic with mock data sets as described in Sec.~\ref{sec:statistics:mock-data-sets}. At large \(\eta\), the LIV modification of the electromagnetic cascade becomes negligible, and the finite-\(\eta\) predictions approach the LI predictions. Consequently, the likelihood-ratio statistic tends to zero. At smaller \(\eta\), the LIV-induced shift of the \(X_{\max,\rm reco}\) distribution and the reconstructed-energy bias make the predictions increasingly incompatible with the observed distributions.

Using the mock-data-set calibration described in Sec.~\ref{sec:statistics:mock-data-sets}, we exclude finite LIV scales for which \(p(\eta)<0.05\). Within the present toy-model treatment, the smooth trend of the calibrated \(p\)-value curve yields the approximate \(95\%\) CL lower limit 
\begin{equation}
M_{\rm LIV}\gtrsim 6.5\times10^{18}\,\text{GeV}.
\label{eq:this-work-constraint}
\end{equation}
The two significant digits in Eq.~\eqref{eq:this-work-constraint} report the rounded position of the smoothing-spline crossing for the adopted interpolation, see Fig.~\ref{fig:lrt-scan}. Given the lack of a first-principles calculation of the transition between the LI and strong-LIV limits in the Bethe--Heitler cross section, this value should be read as an order-of-magnitude estimate of the sensitivity.

At the bound in Eq.~\eqref{eq:this-work-constraint}, photons with \(E_\gamma\gtrsim 3\times10^{17}\,\text{eV}\) are well within the high-energy regime~\eqref{eq:high-energy-regime-for-LIV}. Equation~\eqref{eq:LIV-suppression} gives a Bethe--Heitler cross section of about \(1.4\%\) of its LI value at \(3\times10^{17}\,\text{eV}\), with stronger suppression at higher photon energies. For proton showers of \(10^{19}\,\text{eV}\), at energies relevant to the Auger comparison, the CONEX leading-interaction sample described in Appendix~\ref{app:elm-cascade-development} gives about one to two photons per shower above this energy, carrying together about \(10\%\) of the primary energy on average. The abundance of such photons in the Auger energy range is sufficient for Bethe--Heitler suppression to affect shower development at this LIV scale, providing an independent check on the inferred sensitivity.

As a robustness check, we successively remove the highest reconstructed-energy bins, retaining only bins below \(10^{19}\,\text{eV}\) in the final selection. The smoothed \(p=0.05\) crossing shifts downward overall to \(4.24\times10^{17}\,\text{GeV}\), varying by about one order of magnitude across the cuts, while the calibrated \(p\)-value near the nominal bound stays between \(0.05\) and \(0.09\). The finite Auger bin widths and the \(Y\)- and \(\eta\)-dependent mapping~\eqref{eq:true-energy-from-reco} between reconstructed and true energy prevent this cut from excluding all showers with \(\epsilon>0\), above the simulated energy grid. The reconstructed energy cut removes the showers expected to be most sensitive to LIV, and the observed order-of-magnitude weakening of the bound is consistent with this loss of sensitivity.

It is useful to compare our result with existing constraints on subluminal photon-sector LIV. Bounds at the \(95\%\) CL based on the non-suppression of photon-induced EAS formation reached \(M_{\rm LIV}>1.4\times10^{12}\,\text{GeV}\) from observations of the Crab Nebula by the Tibet-AS\(\gamma\) array~\cite{Satunin2019CrabLIV}, and \(M_{\rm LIV}>1.7\times10^{13}\,\text{GeV}\) from diffuse gamma rays from the Galactic disk observed by the Tibet-AS\(\gamma\) array~\cite{Satunin2021TibetLHAASOLIV}. A~constraint of comparable order, \(M_{\rm LIV}>2.4\times10^{12}\,\text{GeV}\) (\(95\%\) CL), was also obtained from the propagation of TeV gamma rays from astrophysical sources~\cite{Lang2019GammaRayLIV}. A stronger constraint, \(M_{\rm LIV}>2.4\times10^{14}\,\text{GeV}\) (\(95\%\) CL), was obtained in Ref.~\cite{Martynenko2024MuonContentLIV} from the muon content of nuclei-induced EASs.

Further bounds have been obtained under assumptions about the composition and propagation of ultrahigh-energy cosmic rays. Auger obtained \(M_{\rm LIV}\gtrsim 10^{20}\,\text{GeV}\) by comparing upper limits on the ultrahigh-energy photon flux with predictions for photons produced during cosmic-ray propagation in a scenario with a subdominant high-energy proton component~\cite{Auger2022AstroparticleLIV}, and the survival of ultra-high energy cosmic rays implies \(M_{\rm LIV}\gtrsim10^{22}\,\text{GeV}\) assuming proton kinematics and rapid vacuum Cherenkov losses above threshold~\cite{GagnonMoore2004UHECRLIV}.

Constraint~\eqref{eq:this-work-constraint} is numerically much stronger than the bounds quoted above from TeV gamma-ray observations and from the muon content of air showers and is comparable to the prospective constraints expected from the detection of photon-induced EASs with energies around \(10^{17}\,\text{eV}\)~\cite{Rubtsov2013UHEPhotonLIV}. The propagation bounds quoted above rely on either a specified source composition or proton kinematics for the highest-energy primaries. Our shower-based bound is obtained by fitting the mass fractions independently in each reconstructed-energy bin. The strong numerical sensitivity comes from using the full Auger \(X_{\max,\rm reco}\) distributions together with the reconstructed-energy bias induced by the LIV-modified electromagnetic cascade. The interpretation of this constraint in the context of the simplified toy-model treatment is discussed below.

\subsection{Limitations of the toy-model treatment}
\label{sec:results:limitations}

The bound~\eqref{eq:this-work-constraint} should be interpreted in the context of the simplified forward model used in the analysis. The LIV EAS predictions are obtained with a phenomenological modification of the electromagnetic interaction rates in CONEX, rather than with a dedicated detector-level simulation chain built specifically for the LIV scenario. The resulting sensitivity may therefore depend on details of the effective implementation of the LIV-modified electromagnetic cascade.

The simulations are performed for vertical EASs using a single high-energy hadronic interaction model, EPOS~LHC\mbox{--}R. The hadronic-model dependence of the fitted \(X_{\max}\) distributions is not investigated in the present toy analysis.

The detector response is included through the published Auger acceptance and resolution parametrizations rather than through a full detector simulation and event reconstruction. Detector-level effects are included only through these parametrizations. Similarly, the reconstructed-energy response is described by an analytical bias model calibrated on the CONEX samples. This treatment captures the leading effects needed to compare the LIV-modified predictions with the measured \(X_{\max,\rm reco}\) distributions, but accounts only approximately for correlations among reconstructed energy, \(X_{\max,\rm reco}\), event selection, profile-quality cuts, and detector exposure. These limitations are expected to be relevant for a detector-level analysis but are beyond the scope of the present study.

The composition treatment is another important limitation. The fractions of four representative mass groups are fitted simultaneously over all reconstructed-energy bins, and are allowed to vary independently from bin to bin. This flexible treatment reduces the dependence of the LIV constraint on an assumed composition model, but the fitted fractions should not be interpreted as a physical composition measurement. Under the LIV hypothesis, the LIV-induced changes in \(X_{\max,\rm reco}\) and reconstructed energy can modify the mapping between nuclear mass and the observed distributions, so the four-component description may not fully represent a richer true nuclear mixture. In addition, no smoothness constraint is imposed between neighboring reconstructed-energy bins, following the Auger composition-fit strategy.
 
The subluminal photon dispersion in Eq.~\eqref{eq:photon-liv-dispersion} allows vacuum Cherenkov emission for protons with energies \(E_{p}\gtrsim(27m_p^2M_{\rm LIV}^2/4)^{1/4}\)~\cite{Jacobson2002ThresholdEffects}. At our shower-based bound~\eqref{eq:mliv-limit} this threshold corresponds to \(\epsilon=-0.4\). Efficient energy losses during propagation would suppress the proton contribution to the observed \(X_{\max,\rm reco}\) distributions at high energies. The survival of ultrahigh-energy cosmic rays can yield stronger lower bounds on \(M_{\rm LIV}\) under assumptions about their masses, energies, and propagation~\cite{GagnonMoore2004UHECRLIV,Auger2022AstroparticleLIV}. Our analysis provides a complementary estimate from shower development, with the proton fraction fitted freely in each reconstructed-energy bin. A joint analysis should use the same \(M_{\rm LIV}\) to model vacuum Cherenkov losses during propagation and to predict the reconstructed energy and \(X_{\max,\rm reco}\) distributions for each primary mass and true energy. This would test whether the mass composition and primary energies inferred from the shower data are compatible with survival during propagation.

We assume that the electron-sector scale is sufficiently larger than the photon-sector scale. Extending the analysis to arbitrary values of the allowed electron LIV would require a dedicated study. In particular, the soft vacuum Cherenkov process, in which a LIV electron emits a low-energy photon, would need to be included in the EAS simulations. Indeed, the lower bound on the photon-sector LIV scale in Eq.~\eqref{eq:this-work-constraint} is about \(2.5\) orders of magnitude higher than the lower bound on the electron-sector scale in Eq.~\eqref{eq:crab-constraint}.

The vacuum Cherenkov process mainly transfers energy from high-energy electrons and positrons to photons and is not expected to artificially enhance the ordinary electromagnetic cascade. In turn, it may introduce an additional energy-loss channel and could further modify or suppress the electromagnetic cascade. For a more precise bound, this process should therefore be included consistently in the LIV-modified EAS simulation.

Consequently, obtaining a detector-level experimental constraint corresponding to Eq.~\eqref{eq:this-work-constraint} would require a dedicated analysis with a complete treatment of the LIV-modified electromagnetic cascade, full detector simulation and event reconstruction, and comprehensive systematic-uncertainty evaluation.

\subsection{Possible extensions}
\label{sec:results:extensions}

The present analysis uses only the Auger fluorescence-detector \(X_{\max,\rm reco}\) distributions at ultra-high energies. A natural extension would be to repeat the study with event-level simulations and detector reconstruction. This would allow the LIV-modified shower predictions to be propagated through realistic exposure, atmospheric conditions, trigger and selection effects, profile-quality cuts, and correlations between reconstructed energy and \(X_{\max,\rm reco}\). Such an analysis would be needed to turn the shower-maximum-based estimate presented here into a precise experimental constraint.

Another extension would be to include more differential information from the longitudinal profile. The present fit uses only the reconstructed \(X_{\max,\rm reco}\) distributions, while the LIV scenario can also modify the profile shape and the calorimetric-energy response. In this scenario, the effect on the longitudinal profile is not a simple monotonic shift of \(X_{\max}\), because delayed electromagnetic multiplication, reduced effective cascade energy, and late secondary activity can compete. Profile-shape information could therefore help to separate LIV-induced EAS modifications from ordinary composition changes and hadronic-model uncertainties.

It would also be useful to test the method with complementary EAS measurements, including lower-energy fluorescence and Cherenkov-light observations. Although the LIV effect decreases with decreasing primary energy, such measurements provide different detector systematics, different composition sensitivity, and additional information on the longitudinal profile. They could therefore help validate the modeling of LIV-induced electromagnetic-cascade distortions and clarify whether the strong shower-maximum-based sensitivity found here is robust. In particular, the same approach can be applied to the Telescope Array Low-energy Extension~\cite{TelescopeArray2026TALEHybridXmax}, whose published data do not yet provide the level of detail required for the present analysis.

Finally, future studies should extend the physics model of the LIV-modified cascade. In particular, vacuum Cherenkov emission by charged leptons is neglected in the present toy treatment. This omission is acceptable for the present sensitivity estimate, but the process could provide an additional energy-loss channel for high-energy electrons and positrons and may further modify the electromagnetic cascade. For a precise bound, this process should be included consistently in the LIV-modified EAS simulation.

\section{Conclusions}
\label{sec:conclusions}

We present a simulation-based toy analysis of subluminal photon-sector LIV using the \(X_{\max,\rm reco}\) distributions measured by the Auger fluorescence detector. LIV-modified CONEX simulations are used to construct analytical distributions for \(X_{\max,\rm reco}\) and the reconstructed-energy response. These distributions are forward folded with the Auger acceptance and resolution parametrizations and compared with the binned \(X_{\max,\rm reco}\) distributions using a likelihood-ratio test calibrated with mock data sets.

Within this framework, the comparison gives the approximate \(95\%\) CL bound 
\begin{equation}
    M_{\rm LIV}\gtrsim 6.5\times10^{18}\,\text{GeV}.
    \label{eq:mliv-limit}
\end{equation}
The result demonstrates that fluorescence-detector measurements of longitudinal EAS development can provide strong sensitivity to subluminal photon-sector LIV in hadron-induced EASs. This sensitivity arises from the combined effect of LIV-induced modifications of the \(X_{\max,\rm reco}\) distribution and the reconstructed-energy bias induced by the modified electromagnetic cascade.

At the same time, Eq.~\eqref{eq:mliv-limit} should not be read as a final experimental bound from a full detector-level analysis. The present treatment uses simplified detector-response parametrizations, fits the composition with four representative mass groups, and does not include the full experimental reconstruction and systematic-uncertainty chain. A dedicated experimental analysis would require LIV-modified shower predictions propagated through full detector simulation and reconstruction, together with a more complete treatment of LIV in the electron sector and of detector, composition, hadronic-model, and LIV-cascade systematics. The approach presented here thus opens the way for further precise, high-statistics tests of LIV near the Planck scale.

\section*{Acknowledgments}
We are indebted to L.\,A.~Kuzmichev for stimulating discussions that initiated this line of research.

This work was supported by the Russian Science Foundation, grant 22-12-00253(P).

NM and AS thank the Theoretical Physics and Mathematics Advancement Foundation “BASIS” for the student fellowships under the contracts 24-2-10-39-1 and 24-2-10-33-1, respectively. 

The authors used publicly available large language models to assist with English-language editing and selected technical coding tasks, including implementing established algorithms and reviewing source code. The tools were not used for scientific interpretation or for generating the reported results.

We thank the anonymous referee for comments that helped improve the manuscript.

\section*{Data availability}

The preprocessed CONEX shower tables, forward-folded probability tables, measured-count tables, and numerical scan values used in this work are available in the public repository~\cite{GitHubRepo2026AugerFDLIV}.

\appendix

\appsection{Development of the  electromagnetic cascade with Lorentz invariance violation}
\label{app:elm-cascade-development}
We first consider the electrons and positrons produced by Bethe--Heitler conversion of photons in the air shower, assuming subluminal photon-sector LIV and LI electrons and positrons. In this case, energy conservation requires \(E_e<E_\gamma\). An electron (positron) can therefore undergo vacuum Cherenkov emission only if its parent photon has an energy above the lepton threshold in Eq.~\eqref{eq:electron-scale}. This threshold is a factor \((27/8)^{1/4}\simeq1.36\) higher than the characteristic photon energy \((2m_e^2 M_{\rm LIV}^2)^{1/4}\) for Bethe--Heitler suppression.

When Bethe--Heitler pair production is strongly suppressed, only a tiny fraction of photons convert early enough for the resulting pairs to develop a cascade within the available slant depth. Possible effects of hadron-sector LIV on photonuclear interactions are beyond the scope of this work, which assumes LIV only in the photon sector.

In the weak-suppression limit, the vacuum Cherenkov emission is kinematically forbidden, and the bremsstrahlung photons emitted by the daughter electrons (positrons) carry even less energy than their grandparent photons. We therefore expect small photon-sector LIV corrections to their emission.

In the transition region, pair production can supply electrons and positrons for which vacuum Cherenkov emission or modified bremsstrahlung would affect the energy-loss length. The probability of producing such pairs falls rapidly toward the high-energy asymptote. The phenomenological interpolation of the pair-production cross section already introduces uncertainty in the conversion depth in this energy range. We use the standard bremsstrahlung rate and omit vacuum Cherenkov emission in the toy analysis. 

Both approximations affect the predicted longitudinal profile through the shower particles whose energies fall in the transition region. Including vacuum Cherenkov and bremsstrahlung corrections for electrons in this energy interval would be inconsistent with our treatment of comparable LIV effects in the same interval, where we use a smooth phenomenological interpolation to join the cross sections. Particle energies span a broad range within a single shower, and the steep \(\propto E^{-4}_{\gamma}\) dependence in Eq.~\eqref{eq:LIV-suppression} makes the transition region narrow, reducing the probability that a given particle falls within it. 

A more detailed analysis would require a more accurate treatment of the transition region, including a first-principles calculation of how the cross sections connect across it.

To estimate the supply of energetic electrons and positrons from the hadronic cascade, we use a CONEX leading-interaction sample generated with photon--air interactions disabled. The sample is available in the repository accompanying this work~\cite{GitHubRepo2026AugerFDLIV}. It contains \(73\,728\) proton showers at nine energies from \(10^{15}\) to \(10^{19}\,\text{eV}\). Across this range, the mean energy carried by the recorded electrons and positrons is \((3.2\dots 6.0)\times10^{-4}\) of the primary energy, compared with about \(0.29\) for photons, including those from neutral-pion decays. The mean number of recorded electrons and positrons above \(10^{-2}\) of the primary energy is of order \(10^{-2}\) per shower. For a \(10^{19}\,\text{eV}\) proton this cut is close to the vacuum Cherenkov threshold of \(9.3\times10^{16}\,\text{eV}\) at \(M_{\rm LIV}=6.5\times10^{18}\,\text{GeV}\). The leading interactions therefore indicate that the hadronic cascade supplies few leptons energetic enough to radiate near the inferred sensitivity scale.

Electron-sector LIV changes this kinematic argument. Depending on the electron dispersion relation, vacuum Cherenkov emission of soft photons can become allowed, and pair production and bremsstrahlung can also be modified~\cite{Jacobson2002ThresholdEffects,Rubtsov2012LIVCrossSections}. These effects depend on the electron-sector parameters, which are held at their LI values in the present simulations.
\par

\appsection{Profile reconstruction}
\label{app:profile-reconstruction}

The input CONEX profiles are neither fitted as exact curves nor interpreted as detector-level photon counts. Instead, for each simulated EAS, one effective profile realization is constructed by replacing each bin-centered CONEX value of \(\dd E_{\rm dep}/\dd X\) with a random value drawn from a Gaussian distribution centered on that CONEX value. The same variance is assigned to all depth points of a given EAS,
\begin{equation}
    \mathrm{Var}\left[\frac{\dd E_{\rm dep}}{\dd X}\right]
    =
    \left[
        0.2\,
        \max_d
        \left(
            \frac{\dd E_{\rm dep}}{\dd X}
        \right)_d
    \right]^2,
    \label{eq:profile-fit-variance}
\end{equation}
where \(d\) labels the slant-depth bins and the maximum is taken over the bin-centered values of the original CONEX profile for that EAS. The Gaisser--Hillas fit is then applied to this fluctuated profile realization using the variance in Eq.~\eqref{eq:profile-fit-variance}. These fluctuations are introduced to represent a typical scale of fluorescence-detector profile uncertainties at the level needed for the present toy reconstruction.

The regularized likelihood maximized for each profile is defined as
\begin{equation}
    {\cal L}_{\rm GH,reg}
    \equiv
    {\cal L}_{\rm GH}(E_{\rm cal},X_{\max},R,L)\,
    G_{\exp}(L,E_{\rm cal})\,G(R).
    \label{eq:gh-reg-likelihood}
\end{equation}
Here \({\cal L}_{\rm GH}\) is the Gaussian likelihood, while \(G_{\exp}(L,E_{\rm cal})\) and \(G(R)\) are the regularization terms parametrized as in Ref.~\cite{Bellido2023LongitudinalProfiles}.

The explicit formula for \({\cal L}_{\rm GH}(E_{\rm cal},X_{\max},R,L)\) is
\begin{widetext}
\begin{equation}
    {\cal L}_{\rm GH}(E_{\rm cal},X_{\max},R,L)
    \equiv
    \exp\left[
        -\frac{1}{2}
        \sum_d
        \frac{
            \left[
                f_{\rm GH}(X_d \mid E_{\rm cal},X_{\max},R,L)
                -
                \left(\dd E_{\rm dep}/\dd X\right)_d
            \right]^2
        }
        {
            \mathrm{Var}\left[\dd E_{\rm dep}/\dd X\right]
        }
    \right],
    \label{eq:gh-likelihood}
\end{equation}
\end{widetext}
where \(X_d\) is the bin-centered slant depth in bin \(d\), \(\left(\dd E_{\rm dep}/\dd X\right)_d\) is the fluctuated deposited-energy value used in the fit, and the Gaisser--Hillas profile \(f_{\rm GH}\) is written as
\begin{widetext}
\begin{equation}
    \begin{aligned}
    f_{\rm GH}(X\mid E_{\rm cal},X_{\max},R,L)
    =\left(\frac{\dd E_{\rm dep}}{\dd X}\right)_{\max}
    \left[
        1+\frac{R(X-X_{\max})}{L}
    \right]^{R^{-2}}
    \exp\left(
        -\frac{X-X_{\max}}{R L}
    \right),
    \end{aligned}
    \label{eq:gh-profile}
\end{equation}
\end{widetext}
where \(R\) and \(L\) are shape parameters that have been shown to be more independent and stable than the ordinary effective-interaction-scale and first-interaction-depth parameters of the classic Gaisser--Hillas profile parametrization~\cite{Andringa2011LongitudinalProfiles}. The peak energy deposit can be expressed as~\cite{Unger2008ProfileReconstruction}
\begin{equation}
    \left(\frac{\dd E_{\rm dep}}{\dd X}\right)_{\max}=\frac{E_{\rm cal}}{L\,R^{1+2R^{-2}}\,\exp\left(R^{-2}\right)
    \Gamma\!\left(1+R^{-2}\right)},
\end{equation}
where \(\Gamma\) is the gamma function. In this form, the calorimetric energy \(E_{\rm cal}\) acts as the normalization of the fitted profile.

The regularization of \(L\) follows the exponentially modified Gaussian form used in the constrained Auger profile fits. Its location parameter depends on the calorimetric energy through
\begin{equation}
    \langle L\rangle(E_{\rm cal})
    =
    \left[
        227.3
        +
        7.44\,
        \log_{10}
        \left(
            \frac{E_{\rm cal}}{10^{18}\,\text{eV}}
        \right)
    \right]\,\text{g}\,\text{cm}^{-2},
    \label{eq:l-regularization-mean}
\end{equation}
with width \(\sigma_L=11.5\,\text{g}\,\text{cm}^{-2}\) and exponential-tail parameter \(\tau_L=3\sigma_L\)~\cite{Bellido2023LongitudinalProfiles}.

The explicit formula for the \(L\)-regularization term is
\begin{equation}
    \begin{aligned}
    &G_{\exp}(L,E_{\rm cal})
    =
    \frac{1}{2\tau_L}
    \exp\left[
        \frac{\langle L\rangle(E_{\rm cal})-L}{\tau_L}
        +
        \frac{\sigma_L^2}{2\tau_L^2}
        -1
    \right]\\
    &\times
    \operatorname{erfc}
    \left[
        \frac{\langle L\rangle(E_{\rm cal})-L}{\sqrt{2}\sigma_L}
        +
        \frac{1}{\sqrt{2}}
        \left(
            \frac{\sigma_L}{\tau_L}
            -
            \frac{\tau_L}{\sigma_L}
        \right)
    \right].
    \label{eq:l-regularization}
    \end{aligned}
\end{equation}
The regularization of \(R\) is Gaussian,
\begin{equation}
    G(R)
    =
    \exp\left[
        -\frac{1}{2}
        \left(
            \frac{\langle R\rangle-R}{\sigma_R}
        \right)^2
    \right],
    \label{eq:r-regularization}
\end{equation}
where \(\langle R\rangle = 0.257\) and \(\sigma_R = 0.055\)~\cite{Bellido2023LongitudinalProfiles}.

These regularization terms stabilize the profile fit in a way similar to the experimental reconstruction~\cite{Bellido2023LongitudinalProfiles}, while still leaving the profile shape free to respond to the LIV-induced modifications.

The minimization of the negative log-likelihood corresponding to Eq.~\eqref{eq:gh-reg-likelihood} is performed with the Sequential Least Squares Programming (SLSQP) algorithm using analytic gradients. The fitted parameters are constrained to remain positive, and only converged fits are retained.

\begin{figure}[htb]
    \centering
    \includegraphics[width=\linewidth]{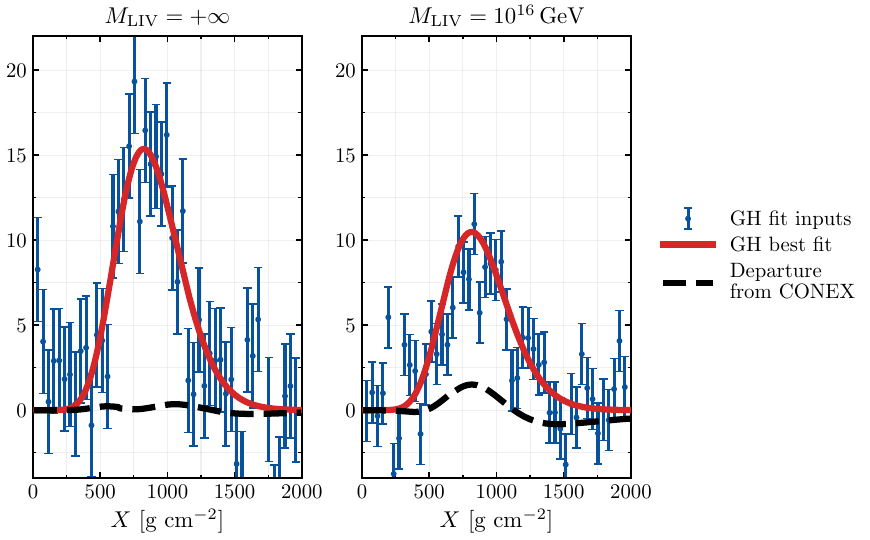}
    \caption{Examples of Gaisser--Hillas profile reconstruction for \({}^{1}{\rm H}\)-induced EASs with primary energy \(E=10^{19}\,\text{eV}\) in the LI case (left) and the LIV case (right). Blue points with error bars represent the fluctuated profile realizations used in the fits. The fluctuated data set is downsampled for illustrative clarity. Solid red curves correspond to the best-fit profiles. Dashed black curves show the deviations of the best-fit profiles from the corresponding CONEX bin-centered profiles.}
    \label{fig:app:GH-fit-examples}
\end{figure}

Figure~\ref{fig:app:GH-fit-examples} demonstrates the fluctuated profile inputs, best-fit Gaisser--Hillas curves, and their deviations from the original CONEX bin-centered profiles for two individual simulated EASs.

\appsection{Shower-maximum distribution parametrization}
\label{app:xmax-parametrization}

The generalized Gumbel density used in Sec.~\ref{sec:model:xmax} is written as
\begin{equation}
    {\cal G}(X_{\max}\mid\epsilon,Y,\eta)
    =
    \frac{1}{\sigma}
    \frac{\lambda^\lambda}{\Gamma(\lambda)}
    \exp\left[-\lambda\left(z+e^{-z}\right)\right],
    \label{eq:xmax-gumbel-density}
\end{equation}
where \(z\equiv\left(X_{\max}-\mu\right)/\sigma\). This form follows Ref.~\cite{DeDomenico2013EASDevelopment}. The parameters \(\lambda\), \(\mu\), and \(\sigma\) control the shape, location, and width of the distribution, respectively. In particular,
\begin{equation}
\begin{aligned}
    \mathrm{E}_{\cal G}\left[X_{\max}\right]
    &=
    \mu
    +
    \sigma
    \left[
        \ln \lambda
        -
        \frac{\dd}{\dd \lambda}\ln\Gamma(\lambda)
    \right],
    \\
    \mathrm{Var}_{\cal G}\left[X_{\max}\right]
    &=
    \sigma^2
    \frac{\dd^2}{\dd \lambda^2}\ln\Gamma(\lambda).
\end{aligned}
\end{equation}

The LI-limit dependence of the generalized-Gumbel parameters on primary energy and mass follows Ref.~\cite{Evoli2026XmaxMoments}. It is written as
\begin{equation}
    \lambda_{\rm LI}(\epsilon,Y)
    =
    \begin{bmatrix}
        1 & \epsilon
    \end{bmatrix}
    W_\lambda
    \begin{bmatrix}
        1 \\ Y \\ Y^2
    \end{bmatrix},
    \label{eq:xmax-lambda-li}
\end{equation}
\begin{equation}
    \mu_{\rm LI}(\epsilon,Y)
    =
    \begin{bmatrix}
        1 & \epsilon & \epsilon^2
    \end{bmatrix}
    W_\mu
    \begin{bmatrix}
        1 \\ Y \\ Y^2
    \end{bmatrix},
    \label{eq:xmax-mu-li}
\end{equation}
and
\begin{equation}
    \sigma_{\rm LI}(\epsilon,Y)
    =
    \begin{bmatrix}
        1 & \epsilon
    \end{bmatrix}
    W_\sigma
    \begin{bmatrix}
        1 \\ Y \\ Y^2
    \end{bmatrix},
    \label{eq:xmax-sigma-li}
\end{equation}
where \(W_{\lambda}\), \(W_{\mu}\), and \(W_{\sigma}\) are coefficient matrices of the LI generalized-Gumbel parametrization.

In the LIV extension used here, the shape and width parameters are kept equal to their LI-limit values at the same \((\epsilon,Y)\),
\begin{equation}
    \lambda(\epsilon,Y,\eta)
    =
    \lambda_{\rm LI}(\epsilon,Y),
    \qquad
    \sigma(\epsilon,Y,\eta)
    =
    \sigma_{\rm LI}(\epsilon,Y),
    \label{eq:xmax-lambda-sigma-liv}
\end{equation}
while the location parameter is shifted according to Eq.~\eqref{eq:xmax-mu-liv}.

The fitted coefficient matrices entering the LI-limit part of the parametrization are given in Table~\ref{tab:xmax-li-coefficients}. The rows correspond to powers of \(\epsilon\), while the columns correspond to powers of \(Y\).

\begin{table}
\caption{Coefficients of the LI-limit part of the generalized-Gumbel parametrization of the \(X_{\max}\) distribution. Each row gives the coefficients multiplying the mass basis \((1,Y,Y^2)\) for a fixed matrix and energy-basis component. The entries of \(W_\mu\) and \(W_\sigma\) are given in \(\text{g}\,\text{cm}^{-2}\), while the entries of \(W_\lambda\) are dimensionless. }
\label{tab:xmax-li-coefficients}
\begin{ruledtabular}
\begin{tabular}{c||c|ccc}
 &  & \(1\) & \(Y\) & \(Y^2\)\\
\hline
\multirow{2}{*}{\(W_\lambda\)}
& \(1\) & \(7.59\times10^{-1}\) & \(5.22\times10^{-1}\) & \(-2.24\times10^{-2}\)\\
& \(\epsilon\) & \(7.28\times10^{-2}\) & \(5.76\times10^{-2}\) & \(-3.14\times10^{-2}\)\\
\hline
\multirow{3}{*}{\(W_\mu\)}
& \(1\) & \(7.93\times10^{2}\) & \(-1.10\times10^{1}\) & \(-1.53\)\\
& \(\epsilon\) & \(5.48\times10^{1}\) & \(-2.06\) & \(7.15\times 10^{-1}\)\\
& \(\epsilon^2\) & \(-2.36\) & \(-4.96\times10^{-1}\) & \(8.09\times10^{-2}\)\\
\hline
\multirow{2}{*}{\(W_\sigma\)}
& \(1\) & \(4.02\times10^{1}\) & \(\ \ 4.30\) & \(-1.59\)\\
& \(\epsilon\) & \(-7.74\times 10^{-1}\) & \(-1.48\) & \(1.26\times10^{-1}\)\\
\end{tabular}
\end{ruledtabular}
\end{table}

The parametrization contains 26 fitted parameters in total: the matrices \(W_\lambda\), \(W_\mu\), and \(W_\sigma\), together with the five parameters of Eq.~\eqref{eq:xmax-mu-shift}. The fitted covariance matrix of the 26-parameter model is propagated only in the mock-data-set calibration described in Sec.~\ref{sec:statistics:mock-data-sets}.

\appsection{Auger detector response}
\label{app:auger-response}

The forward folding described in Sec.~\ref{sec:forward-folding:detector} uses the fluorescence-detector-response parametrizations provided with the Auger reconstructed-shower-maximum distributions~\cite{Auger2026XmaxFD,Auger2026XmaxDataset}.

For each reconstructed-energy bin \(b\), the acceptance is described by four tabulated parameters, \(X_{1,b}\), \(X_{2,b}\), \(L_{1,b}\), and \(L_{2,b}\). We write the acceptance as
\begin{equation}
    {\cal A}_{b}(X_{\max})=
    \left\{
    \begin{array}{@{}l@{\quad}l@{}}
        \exp\left(\dfrac{X_{\max}-X_{1,b}}{L_{1,b}}\right), & X_{\max}<X_{1,b},\\[0.4em]
        1, & \makebox[2.1cm][r]{\text{\(X_{1,b}\le X_{\max}\le X_{2,b},\)}}\\[0.4em]
        \exp\left(\dfrac{X_{2,b}-X_{\max}}{L_{2,b}}\right), & X_{\max}>X_{2,b}.
    \end{array}
    \right.
    \label{eq:xmax-acceptance}
\end{equation}

The detector resolution in \(X_{\max}\) is represented by a Gaussian kernel,
\begin{equation}
    \begin{aligned}
    &{\cal R}_{b}(X_{\max,\rm reco}-X_{\max})
    =\\
    &=
    \frac{1}{\sqrt{2\pi} \varsigma_{b}}
    \exp\left[
        -\frac{\left(X_{\max,\rm reco}-X_{\max}\right)^2}{2 \varsigma_{b}^2}
    \right],
    \end{aligned}
    \label{eq:xmax-resolution-kernel}
\end{equation}
where \(\varsigma_b\) is the tabulated effective resolution width for reconstructed-energy bin \(b\).

\appsection{Statistical procedure details}
\label{app:statistical-procedure}

For fixed \(\eta\) and \(F\), the expected number of events in bin \((x,b)\) is \(\nu_{x,b}(\eta,\mathbf{f}_b)\), defined in Eq.~\eqref{eq:expected-counts}. The ordinary Poisson likelihood used to construct the likelihood-ratio statistic is
\begin{equation}
    {\cal L}(\eta,F)
    =
    \prod_b\prod_x
    \frac{
        \nu_{x,b}(\eta,\mathbf{f}_b)^{n_{x,b}}
        \exp[-\nu_{x,b}(\eta,\mathbf{f}_b)]
    }{
        \Gamma(n_{x,b}+1)
    }.
    \label{eq:poisson-likelihood}
\end{equation}

Following the Auger composition-fit convention~\cite{Auger2014XmaxComposition}, the composition fractions are fitted with a likelihood ratio relative to the saturated model, for which the predicted bin counts are equal to the observed counts. With the reconstructed-energy-bin weights used in this work, the quantity minimized in the fit is
\begin{widetext}
\begin{equation}
    -\ln{\cal L}'(\eta,F)
    =
    \sum_b \omega_b
    \sum_x
    \left[
        \nu_{x,b}(\eta,\mathbf{f}_b)
        -
        n_{x,b}
        +
        n_{x,b}
        \ln
        \left(
            \frac{n_{x,b}}
                 {\nu_{x,b}(\eta,\mathbf{f}_b)}
        \right)
    \right],
    \label{eq:weighted-fit-objective}
\end{equation}
\end{widetext}
where the logarithmic term is defined to be zero when \(n_{x,b}=0\). The factors \(\omega_b\equiv N_{b,\rm obs}^{-1}\) are fixed weights assigned to reconstructed-energy bins. They put the contributions from different reconstructed-energy bins on a comparable numerical scale in the simultaneous implementation. They do not couple the composition fractions between different reconstructed-energy bins. This choice follows the Auger composition analysis, where the reconstructed-energy bins are fitted separately; here all bins are included in one simultaneous fit for implementation convenience.

For each tested LIV scale, the fitted composition fractions are defined by the constrained minimization
\begin{equation}
    \widehat F(\eta)
    \equiv
    \underset{F}{\operatorname{arg\,min}}\,
    \left[-\ln{\cal L}'(\eta,F)\right],
    \label{eq:fitted-composition}
\end{equation}
where the minimization is performed over \(F\) under the positivity and normalization constraints in Eq.~\eqref{eq:composition-simplex}. Because the composition vectors are independent in different reconstructed-energy bins, this constrained minimization is separable over \(b\) for fixed \(\eta\). The constrained minimization is performed with the SLSQP algorithm using analytic gradients. To reduce sensitivity to boundary configurations and local minima, the minimization is repeated from \(10\) independent initial composition arrays. Each array is constructed by drawing one composition vector from the \(\operatorname{Dirichlet}(1,1,1,1)\) distribution in every reconstructed-energy bin. The solution with the smallest value of the weighted objective in Eq.~\eqref{eq:weighted-fit-objective} is retained.

The value entering Eq.~\eqref{eq:lrt-statistic} is the ordinary Poisson log-likelihood evaluated at the fitted fractions,
\begin{equation}
    \ln{\cal L}_\eta
    \equiv
    \ln{\cal L}(\eta,\widehat F(\eta)).
    \label{eq:fitted-poisson-logl}
\end{equation}
Thus the weighted likelihood-ratio expression in Eq.~\eqref{eq:weighted-fit-objective} is used to determine the composition fractions, while the final statistic \(D(\eta)\), defined in Eq.~\eqref{eq:lrt-statistic}, is constructed from the unweighted Poisson likelihood.

For a fixed finite value of \(\eta\), mock data sets are generated from the fitted LIV prediction at that value of \(\eta\). In each mock data set, the single-mass-group probabilities \({\cal P}_{x,b,y}(\eta)\) are first fluctuated according to their total estimated uncertainties. The fluctuations are sampled independently for different \(X_{\max,\rm reco}\) bins, reconstructed-energy bins, and mass groups, using Gaussian distributions centered on the nominal probabilities. Negative fluctuated probabilities are clipped to a small positive value, and the resulting distributions are renormalized separately for each reconstructed-energy bin and mass group. The fluctuated distributions are then mixed with the fitted composition fractions and normalized to the observed number of events in each reconstructed-energy bin, as in Eq.~\eqref{eq:expected-counts}. The fluctuated distributions are used only to generate mock counts; the subsequent fits use the nominal distributions, as in the fit to the observed data.

Mock counts \(n_{x,b,i}\) are generated independently in each bin according to
\begin{equation}
    n_{x,b,i}
    \sim
    {\rm Poisson}
    \left[
        \nu_{x,b,i}(\eta,\widehat F(\eta))
    \right].
    \label{eq:mock-poisson}
\end{equation}
Here \(i\) labels the mock data set, and \(\nu_{x,b,i}\) denotes the expected counts obtained after the probability perturbation used in that mock data set.

For every mock data set, the finite-\(\eta\) and LI hypotheses are refitted using the same fitting procedure as for the observed data. For mock data set \(i\), this gives
\begin{equation}
    D_i(\eta)
    =
    -2\left(
        \ln{\cal L}_{\eta,i}
        -
        \ln{\cal L}_{+\infty,i}
    \right),
    \label{eq:mock-lrt}
\end{equation}
where \(\ln{\cal L}_{\eta,i}\) and \(\ln{\cal L}_{+\infty,i}\) are the unweighted Poisson log-likelihoods after composition fitting for mock data set \(i\) under the finite-\(\eta\) and LI hypotheses, respectively.

We estimate the additional spread of the mock statistic \(D_i(\eta)\)
associated with discrepancies between the analytical shower
parametrization and the CONEX distributions. Consider a perturbation
\(n_{x,b,i}\to n_{x,b,i}+\delta n_{x,b,i}\) of mock data set \(i\).
Holding the fitted composition fractions fixed, the first-order change in Eq.~\eqref{eq:mock-lrt} is
\begin{equation}
    \delta D_i(\eta)
    =
    -2\sum_{x,b}\delta n_{x,b,i}
    \frac{\partial}{\partial n_{x,b,i}}
    \left(\ln{\cal L}_{\eta,i}
          -\ln{\cal L}_{+\infty,i}\right).
    \label{eq:mock-lrt-perturbation-definition}
\end{equation}
The count-factorial terms cancel between the two Poisson
log-likelihoods. Consequently,
\begin{equation}
    \delta D_i(\eta)
    =
    2\sum_{x,b}\delta n_{x,b,i}
    \ln
    \frac{
        \nu_{x,b}(+\infty,\widehat{\mathbf{f}}_{b,i}(+\infty))
    }{
        \nu_{x,b}(\eta,\widehat{\mathbf{f}}_{b,i}(\eta))
    },
    \label{eq:mock-lrt-perturbation-expression}
\end{equation}
where \(\widehat{\mathbf{f}}_{b,i}\) denotes the composition fitted to mock data set \(i\). Thus a given uncertainty in the mock counts has a scale-dependent effect on \(D_i(\eta)\), determined by the difference between the fitted LI and LIV predictions.

To estimate the count perturbations, we use the empirical CDF (\({\rm ECDF}\)) of the variable \(u_{\cal G}\) defined by Eq.~\eqref{eq:uniform-variable} at each simulated scale \(\eta_s\) (including LI) and combining showers of all primary energies
and masses. For an exactly described distribution,
\({\rm ECDF}(u_{\cal G})=u_{\cal G}\). Hence, for a bin with cumulative-probability
boundaries \((u_{\cal G})_{x-1,b}\) and \((u_{\cal G})_{x,b}\), the discrepancy in its
probability is
\begin{equation}
    \begin{aligned}
    \delta{\cal P}_{x, b}(\eta_s) \equiv {\rm ECDF}\left((u_{\cal G})_{x,b}\right)-{\rm ECDF}\left((u_{\cal G})_{x-1,b}\right)
    -\\
    -\left((u_{\cal G})_{x,b}-(u_{\cal G})_{x-1,b}\right).
    \end{aligned}
    \label{eq:probability-perturbation}
\end{equation}
We therefore estimate the corresponding count perturbation as
\begin{equation}
    \delta n_{x,b}(\eta_s)
    \equiv
    N_{b,\text{obs}}\,
    \delta {\cal P}_{x, b}(\eta_s).
    \label{eq:count-perturbation}
\end{equation}
The boundaries are taken from a common LI template, used at every tested LIV scale. The first and last cumulative boundaries are fixed to zero and unity, so that \(\sum_x\delta n_{x,b}(\eta_s)=0\). This transfer of the CDF discrepancies to reconstructed bins is an approximation that does not take into account detector-level effects.

Each perturbation \(\delta n_{x,b}(\eta_s)\) defines a possible
shift \(\delta_s D_i(\eta)\) through the first-order
expression above. For the numerical estimate, we replace its mock-dependent logarithmic coefficients by common coefficients: the finite-\(\eta\) prediction uses the saved composition fit to the observed data, and the LI prediction uses the common template described above. With this approximation,
\(\delta_s D_i(\eta)\simeq\delta_s D(\eta)\). We treat the mean squared shift, with equal weights for all simulated scales, as an additional variance contribution. This prescription corresponds to assigning either sign to each estimated shift with equal probability, independently of the original mock fluctuation; it accounts for uncertainty in the parametrization without applying a preferred shift to the mock mean. We implement the resulting variance increase
by the replacement \(D_i(\eta)\) with
\begin{equation}
    \begin{aligned}
    D_{i,\text{corr}}(\eta)&\equiv\left\langle D_i(\eta)\right\rangle_i
    +\\
    &+
    \left[
        1+
        \frac{
            \left\langle
                [\delta_s D (\eta)]^2
            \right\rangle_s
        }{
            \operatorname{Var}_i[D_i(\eta)]
        }
    \right]^{1/2}
    \left[
        D_i(\eta)-\left\langle D_i(\eta)\right\rangle_i
    \right].
    \end{aligned}
    \label{eq:mock-lrt-broadening}
\end{equation}

The calibrated \(p\)-value for the tested LIV scale is estimated as 
\begin{equation}
    p(\eta)
    =
    \frac{
        1+\sum_{i=1}^{N_{\rm mock}}
        \Theta\!\left[
            D_{i,\text{corr}}(\eta)-D_{\rm obs}(\eta)
        \right]
    }{
        1+N_{\rm mock}
    },
    \label{eq:mock-pvalue}
\end{equation}
where \(N_{\rm mock}=10^4\), \(\Theta\) is the Heaviside step function, and \(D_{\rm obs}(\eta)\) is the value obtained from the observed Auger counts (without applying a correction). The \(+1\) correction avoids assigning zero probability to the observed tail when the number of mock data sets is finite.

The public repository~\cite{GitHubRepo2026AugerFDLIV} provides the processed numerical inputs needed to reproduce the statistical scan from the binned-probability level. These include the preprocessed CONEX shower tables, the single-mass-group probabilities with their estimated uncertainties, the reconstructed-energy and \(X_{\max,\rm reco}\) binning, and the measured-count tables. The repository~\cite{GitHubRepo2026AugerFDLIV} also contains the numerical values of the likelihood-ratio statistic and the \(p\)-values obtained from the mock-data-set calibration.

\bibliography{liv-fd}
\end{document}